\documentclass[%
reprint,
 amsmath,amssymb,
 aps,
]{revtex4-2}

\usepackage[final]{changes}
\definechangesauthor[color=black]{msb}
\definechangesauthor[color=red]{m}

\usepackage{comment}
\usepackage{graphicx}
\usepackage{dcolumn}
\usepackage{bm}

\newcommand{\norm}[1]{\left\lVert#1\right\rVert}

\begin{document}

\setlength{\abovedisplayskip}{3pt}
\setlength{\belowdisplayskip}{3pt}


\title{Wonders of chaos for communication}

\author{Murilo S. Baptista}
 \affiliation{Institute for Complex Systems and Mathematical Biology, Department of Physics, University of Aberdeen, AB24 3UX, Aberdeen, UK.}


\date{\today}

\begin{abstract}
\added[id=m]{This work shows that chaotic signals with different power spectrum are robust to linear superposition, meaning that  the superposition preserves Ergodic quantities (Lyapunov exponents) and the information content of the source signals, even after being transmitted over non-ideal physical medium. This wonderful property that chaotic signals have allows me to propose a novel communication system based on chaos, where information composed from and to multiple users each operating with different base frequencies and that is carried by chaotic wavesignals can be fully preserved after transmission in the open air wireless physical medium, and it can be trivially decoded with low probability of errors.} 
\added[id=m]{
This work tackles with great detail how chaotic signals and their information content are affected when travelling through medium that presents the non-ideal 
properties of multipath propagation, noise and chaotic interference (linear superposition), and how this impacts on the proposed communication system. Physical media with other non-ideal properties (dispersion and interference with periodic signals) are also discussed.}
\end{abstract}


\maketitle


\section{Introduction}\label{Introduction}

Communication systems are designed to cope with the constraints of the physical medium. Previous works have shown that 
chaos has intrinsic properties that makes it attractive to sustain the modern design of communication systems. 

\added[id=m]{Three seminal works \cite{hayesPRL1993,cuomo1993circuit,CSK1993a} have launched the area of communication with chaos. 
In Ref.  \cite{hayesPRL1993}, the authors have controlled chaos  \cite{OGY1990}
to propose a communication system where signals would represent desired digital information, yet preserving the original dynamics of the chaotic system. 
In Ref. \cite{cuomo1993circuit}, the authors have proposed a communication system in which analog information signals were added (masked) into the chaotic signal to be transmitted. Synchronisation of chaos between transmitter and the receiver \cite{pecora1990} was used as a means to extracted the information at the receiver end. In Ref. \cite{CSK1993a}, the authors have shown that chaos can be 
modulated to represent arbitrary digital information. Later on, other works have followed \cite{Parlitz1994,Parlitz1995}, 
where in Ref. \cite{Parlitz1994}, it was shown that chaotic signals could be modulated in spread sequences 
to represent arbitrary digital information, and in Ref.
\cite{Parlitz1995}, a general method to communicate based on the synchronization of chaotic systems was proposed. Since then, the area has attracted 
great attention to the scientific community, producing a vast collection of research.}

Take $x(t)$ to  represent a controlled chaotic signal and that encodes information from a single transmitter.  Let $r(t)$ to represent the transformed signal that is received. 
Chaos has offered communication systems whose information capacity could remain invariant by a small increase in the noise level,  \cite{cuomo1993circuit,bollt1997coding,rosa1997noise,baptista2002information}, could be robust to filtering \cite{badii1988dimension,eisencraft2012chaos,ren2013wireless,eisencraft2012chaos}, and multipath propagation \cite{ren2013wireless}, intrinsically present in the wireless communication. Decoding of $r(t)$ can be trivial, 
with the use of a simple threshold technique   \cite{ren2013wireless,ren2016experimental}. 
Chaos allows for simple controlling techniques to encode digital information \cite{hayesPRL1993,bollt2003}. For the wonderful solvable systems proposed in \cite{corron2010matched,corron2015chaos}, simple analytical expressions to generate the 
controlled signal $x(t)$ can be derived \cite{patentUK,ren-experiment2019}. Moreover, these systems have matched filters whose output maximises the signal to noise ratio (SNR) of $r(t)$, thus offering a practical and reliable way to decode transmitted information. 
Chaos allows for integrated communication protocols \cite{baptista2000integrated}, it offers viable solutions for 
the wireless underwater \cite{ren2017chaotic,bai2018chaos}, digital \cite{bailey2016} and optical \cite{Jiang:16} communication, radar applications \cite{corron-radar2016}, and simultaneously radar-communication  \cite{carroll}. Chaotic
communication has been experimentally shown to achieve higher bit rate in a commercial wired fiber-optic channel \cite{argyris2008photonic}, 
and lower Bit-Error-Rate (BER) than conventional wireless non-chaotic baseband waveform methods. Moreover, chaos-based communication only requires equipment that is compatible with the today commonly-used ones \cite{ren-experiment2019}. 

\added[id=m]{In order to embed the present work into the 
state of the art, I propose a classification for works in this area: (i) those methods that control the chaotic signal to represent 
(encode) information, yet preserving relevant invariant properties of chaos, as in Refs. \cite{hayesPRL1993,patentUK,ren-experiment2019}; (ii) those that manipulate/modulate the chaotic signal for it to represent (encode) the desired information by altering some natural property of the signal, for example its amplitude, 
as the chaos shift keying \cite{CSK1993,CSK1993a} and the spreading method of \cite{Parlitz1994}); (iii) those in which information is extracted at the receiver (decoded) by means of visual inspection of the received signals, as in  Refs. \cite{hayesPRL1993,ren2013wireless}; (iv) those in which extraction of information happens via a matched filter as in Ref. \cite{corron2010matched} (also known as coherent matched filter receiver \cite{chua1998}); (v) those in which information is extract by means of synchronization (also known as coherent correlation receiver with chaotic synchronization \cite{chua1998}), as in Ref. \cite{cuomo1993circuit,Parlitz1995}. In the review on digital communication with chaos in Ref. \cite{chua1998}, two additional schemes are discussed, noncoherent detection techniques, and differentially coherent reception. The present work belongs to the 4 first categories ((i)-(iv)). I do assume that there is a control producing a chaotic signal that encodes a desired arbitrary digital message (i), the signal has its amplitude modified by a power gain (ii), decoding is done by a simple visual inspection of the 
received signal (iii), and information can be extracted by means of a matched filter at the receiver (iv). However, my work additionally explore the wonderful decomposability property chaotic signals have that enables a solution for a multi-source and multi-frequency communication, a work that can allow chaos to be adopted as a native signal to support wireless networked communication 
systems such as the Internet of Things or 5G.}

\added[id=m]{Several works on communication with chaos have focussed on a system composed by 2 users, the transmitter and the receiver. Some works were motivated by the master-slave synchronization configuration \cite{pecora1990} where the master (the transmitter) sends the information to the slave (the receiver) \cite{cuomo1993circuit}. The understanding of how 2 users communicate cannot always capture the complexities involved in even simple 
networked communication systems.  
It is often more appropriate to break down this complex communication problem into a much simpler problem consisting of 2 configurations, the uplink and the downlink. The uplink configuration would render us an understanding of how several nodes that transmit different information signals can be processed in a unique central node.  The downlink configuration would render us an understanding about how a unique central node that transmits a single signal can distribute dedicated information for several other nodes. This strategy to break a complex network problem into several smaller networks describing the uplink and the downlink configurations, and that is crucial to understand very complex technologically oriented flow networks, such as the communication and power networks, can also shed much light into the processing of information in networks as complex as the brain. The uplink configuration would contribute to a better understanding about how pre-synaptic neurons transmit information to a hub neuron, and the downlink configuration would contribute to a better understanding about how post-synaptic neurons can process information about a hub neuron. This paper focus on information signals that are linearly composed, and thus, this approach could in principle be used to explain communication in neurons doing electric synapses. However, the main focus of the present paper is about the understanding of how superimposed chaotic signals can be robust to non-ideal properties of physical medium that is present in wireless communication networks.}

In a scenario where the received signal, $r(t)$,  \added[id=msb]{is composed by a linear superposition of chaotic signals} of two transmitters $x^{(1)}$ and $x^{(2)}$ (or more), as in $r(t)=\added[id=msb]{\tilde{\gamma}^{(1)}} x^{(1)}(t) + \added[id=msb]{\tilde{\gamma}^{(2)}}x^{(2)}(t) + w(t)$, each signal operating with different frequency bandwidths and each encoding different information contents, with $\added[id=msb]{\tilde{\gamma}^{(i)}} \in \Re$ and $w(t)$ representing Additive White Gaussian Noise (AWGS) modelling the action of a physical medium in the composed transmitted signal, is it possible to decompose the source signals,  $x^{(1)}(t)$ and 
$x^{(2)}(t)$, out of the received signal $r(t)$, and recover (i.e., decode) their information content? 


In this paper, I show that for the no noise scenario, the spectrum of  positive Lyapunov Exponents (LEs) of $r(t)$ is the union of the set of the positive Lyapunov exponents of both  signals $x^{(1)}(t)$ and $x^{(2)}(t)$. \added[id=msb]{This is demonstrated in the main manuscript} \added[id=m]{in Sec. \ref{preservation-LEs} for the system used to communicate. Appendix C  generalizes this result to superimposed signals coming from arbitrary chaotic systems.} And what is more, for the system proposed in \cite{corron2010matched}, the information content of \added[id=msb]{the composed signal} $r(t)$ preserves the information carried by the source signals, \added[id=msb]{this being linked to the preservation of the positive Lyapunov exponents}.  \added[id=m]{This result is fully explained in Sec. \ref{LE-and-information}, where I present the information encoding capacity of the proposed communication system, or in other words, the rate of information contained in the linearly composed signal of several chaotic sources. I also discuss in this Section how this result extends to communication systems that have users communicating with other chaotic systems, different from the one in Ref. \cite{corron2010matched}.}
\added[id=msb]{Preservation of the Lyapunov exponents in the composed signals of arbitrary chaotic systems is demonstrated thanks to an equivalence principle} deterministic chaotic systems have that permits that the composed signal 
can be effectively described by a signal departing from a single source \added[id=msb]{but with time-delayed components}. 
Moreover, when the physical medium where the composed signal is transmitted has noise, it is possible to determine appropriate linear coefficients \added[id=msb]{$\tilde{\gamma}^{(i)}$} (denoted as power gains, see Eq. (\ref{power-gains}) in Sec. \ref{channel-encoding}), which will depend on the natural frequency of the user, on the attenuation properties of the media and the number of users (end of Sec. \ref{encoding}),  such that the information content carried by the composed signal  $r(t)$ \added[id=msb]{can be trivially decomposed,} or decoded, by a simple threshold (see Eq. (\ref{decoding-partition})), with low-probability of errors, or no errors at all for sufficiently small noise levels. \added[id=m]{In the later case, that would imply that the information encoding capacity provides the information capacity of the system, or the rate of information received/decoded.}

\added[id=msb]{The scientific problem to decompose a linear superposition of chaotic signals that renders the mathematical support for the proposed communication system is similar to that of blind source separation for mixed chaotic signals \cite{Yi2012} or that of the separation of a signal composed of a linear superposition of independent signals \cite{ott2020}. However, these separation methods require long measurements, and additionally either several measurements of multiple linear combinations of the source signals, or source signals that have similar power spectra and that are independent. These requirements cannot be typically fulfilled by a typical wireless communication environment, where information must be decoded even when very few observations are made, signals are sent only once with constant power gains, source signals can have arbitrary natural frequencies, and they can be dependent.} 


  
I also show \added[id=m]{in Sec. \ref{decoding}}
that in the \added[id=msb]{single-user communication system} proposed in the work of  \cite{corron2010matched}, \added[id=msb]{with a chaotic generator for the source signal and a matched filter to decode information from the received signal corrupted by noise, the chaotic generator has no negative LEs, which leads to a stable matched filter with no positive LEs, and that can therefore optimally filter noise. Moreover, I show that the single-user communication system formed by the chaotic generator plus the matched filter} can be roughly approximated by the unfolded-Baker's map \cite{lasota1985probabilistic}. \added[id=msb]{This understanding permits the conclusion that in the multi-user environment the matched filter that decomposes the source signal of a user from the received composed signal $r(t)$ is the matched filter of that user alone.}

\added[id=m]{I will then study, in Sec. \ref{performance},  the information capacity of the proposed communication system in prototyped wireless network configurations, and in Sec. 
\ref{Wi-C1-versus-Noma}, I will compare its performance with a non-chaotic communication method that is the strongest candidate for the future 5G networks, the Non-Orthogonal Multiplex Access (NOMA), and will show that the proposed multi-user chaos based communication system can (under certain configurations) communicate at higher bit rates for large noise levels in the physical medium.}

\added[id=m]{Finally, in Sec. \ref{NL-channels}, I will discuss how communication with chaos can be made robust to other types of non-ideal physical media (also refereed as a ``channel of communication'')\cite{EISENCRAFT} that present dispersion and whose signals interfere with other period (non-chaotic) signals.}

\section{Linear composition of chaotic signals, the preservation of the Lyapunov exponents and encoding for transmission.}\label{encoding}


A wonder of chaotic oscillations for communication is the system proposed in Ref. \cite{corron2010matched}. 
With an appropriate rescaling of time to a new time-frame $dt{^{\prime}}=\gamma dt$, it can be rewritten as 
\begin{equation}
    \ddot{x}-2\beta(\gamma)\dot{x}+(\omega^2+\beta(\gamma)^2)(x-s(t))=0, 
\label{rescaled-corron}
\end{equation}
\noindent 
where $s(t) \in (-1,1)$ is a 2-symbols alphabet discrete state that switches value by the signum function $s(t)={x(t)/|x(t)|}$, whenever $|x(t)|<1$ and $\dot{x}=0$. If the information to be communicated is the binary stream $\textbf{b}=\{b_0,b_1,b_2, \ldots\}$ ($b_n \in \{0,1\}$) a signal can be created such that $s(t) = (2b_n-1)$, for $nT \leq t < (n+1)T$ \cite{patentUK}. In this new time-frame, the natural frequency is $f(\gamma)=1/\gamma$ ($\omega=2\pi f$), the period $T(\gamma)=1/f(\gamma)=\gamma$, and $\beta(\gamma) =\beta(\gamma=1) f(\gamma)$, where $0 < \beta(\gamma=1) \leq \ln{(2)}$. 
More details can be seen in \textbf{Appendix A}. \added[id=m]{$\beta(\gamma=1)$ is a parameter, but with an important physical meaning. It represents the Lyapunov exponent (LE) of the system in units of nepits per period (or per cycles), which is also equal to the rate of information produced by the chaotic trajectory in nepits per period. 
On the other hand, $\beta(\gamma)$ represents the LE in units of nepits per unit of time,  which is also equal to the rate of information produced by the chaotic trajectory in nepits per unit of time.  See Sec. \ref{LE-and-information}.}

The received signal in the noiseless wireless channel from user $k$ can be modelled by  
\begin{equation}
r^{(k)}(t)    = \sum_{l=0}^{L-1} \alpha_l \gamma^{(k)} x(t-\tau_l)
\label{wireless_channel-one-user}
\end{equation}
\noindent
where there are $L$ propagation paths, each with an attenuation factor of $\alpha_l$ and a time-delay $\tau_l$ for the signal to arrive to the receiver along the path $l$ (with $0=\tau_0 < \tau_2 < \ldots < \tau_{L-1}$), and 
$ \gamma^{(k)}$ is an equalising power gain \added[id=m]{to compensate for the amplitude decay due to the attenuation factor}. The 
noisy channel can thus be modelled by $r(t)+w(t)$, where $w(t)$ is an AWGN. 

Let me consider \added[id=msb]{the time-discrete dynamics of the signal generated by a} single user $r^{(k)}(t) = r(t)$ (with $\gamma^{(k)}=1$), whose signal is sampled at frequency $f$, so $r_n=r(n/f)$ are collected, then the return map (\textbf{see Appendix B}) of the received signal (assuming for simplicity that $ \gamma^{(k)}$=1) is given by 
\begin{equation}
  r_{n+1} = e^{\frac{\beta}{f}} r_n - \sum_{l=0}^{L-1}  \alpha_l \left( 
    e^{\beta/f} s_{n^{\prime}} - \mathcal{K}_l s_{n^{\prime}} - 
    s_{n^{\prime}+1} + s_{n^{\prime}+1} \mathcal{K}_l \right)
\label{multipath-returnmap-MSB}
\end{equation}
\noindent
where 
$n^{\prime} = n - \lceil f \tau_l \rceil$ and 
$\mathcal{K}_l = e^{-\beta(\tau_l - \lceil \tau_l/T \rceil T)}[\cos{\left(2\pi\frac{\tau_l}{T}\right)} + \frac{\beta}{\omega}\sin{\left(2\pi \frac{\tau_l}{T}\right)}]$ 
\noindent
where $s_n$ represents the binary symbol associated to the time interval $nT \leq t <(n+1)T$, so $s_n=s(t=nT)$, $\lceil f \tau_l \rceil$ representing the ceiling integer of $f \tau_l$, and $\frac{\beta}{f}$ denotes $ \frac{\beta(\gamma)}{f(\gamma)} = \beta(\gamma=1)$. Equation (\ref{multipath-returnmap-MSB}) extends the result in \cite{yao2017chaos}, valid for when $\tau_l = mT$, with $m \in \mathbb{N}$, when $\mathcal{K}_l=1$.

The \added[id=msb]{Lyapunov exponent} (LE) of the \added[id=msb]{1-dimensional} map in Eq. (\ref{multipath-returnmap-MSB}) in units of nepits per period for multi-path propagation, \added[id=msb]{denoted by} $\chi$, (which is equal to the positive LE of the continuous dynamics - \textbf{see Sec. I of Supplementary Material (SM)}) is equal to \added[id=m]{$\chi = \frac{\beta}{f} = \beta(\gamma=1)$ [nepits per period]}, since $\chi = \lim_{n\to\infty} \frac{1}{n} \ln{\left|  \prod_{i=0}^{n} \frac{dr_{n+1}}{dr_n} \right|}$. \added[id=m]{This LE can be calculated in nepits per unit of time by simply making  $\frac{\chi}{T} = \beta$.} \added[id=m]{LE can be calculated in units of ``bits per period" by using binary logarithm instead of natural logarithm}. This is also equal to the LE of the return map 
\begin{equation}
x_{n+1}=e^{\frac{\beta}{f}}[x_n - (1-e^{-\beta/f})s_n],
\label{returnmap}
\end{equation}
\noindent
obtained from Eq. (\ref{multipath-returnmap-MSB}) when there is only a direct path, $L=1$. Notice also that the constant attenuation factor $\alpha_l$ does not contribute to this LE, only acting on the value of the binary symbols. This is to be expected 
\cite{maria2019}.

\subsection{Linear composition of chaotic signals for the uplink and the downlink communication configurations}

The analysis will focus on two prototype wireless communication configurations: the uplink  and the downlink. In the uplink communication, several users transmit signals \added[id=msb]{that become linearly superimposed when they arrive} to a base station antenna (BS). In the downlink communication, a BS sends 1 \added[id=msb]{composed signal (linear superposition of chaotic signals)} signal containing information to be \added[id=msb]{decomposed (or decoded)} by several users. 

I propose a chaos-based communication system, named "Wi-C1", that allows for multi-user communication, where \added[id=msb]{one of the N users} operates with its own natural frequency. It is assumed other constraints of the wireless medium are present, such as 
multipath propagation and AWGN. Wi-C1 with 1 BS can be modeled by \added[id=msb]{a linear superposition of chaotic signals as}
\begin{equation}
 O(t)_{u}    = \sum_{k=1}^N  \sum_{l=0}^{L^{(k)}-1} \alpha_l^{(k)} \gamma^{(k)} \tilde{\gamma}^{(k)} x^{(k)}(t-\tau_l^{(k)}) + w(t) \label{WiChaos_up}
\end{equation}
\begin{equation}
O^{(m)}(t)_{d}  =  \sum_{l=0}^{L^{(k)}-1} \alpha_l^{(m)} \sum_{k=1}^N \gamma^{(k)} \tilde{\gamma}^{(k)} x^{(k)}(t-\tau_l^{(m)})  +  w^{(m)} \label{WiChaos_down}\nonumber
\end{equation}

$O(t)_{u}$ in Eq. (\ref{WiChaos_up}) represents \added[id=msb]{the composed} signal received at BS from all users in the uplink. \added[id=msb]{This signal will be the focus of the paper from now on}. $O^{(m)}(t)_{d}$ represents the signal received by user $m$ from a \added[id=msb]{composed signal} transmitted by the BS in the downlink. 
$w(t)$ represents an AGWN at the base station, 
and $w^{m}(t)$, for $m=1,\ldots,N$ represents AGWN at the user $m$. $\alpha_l^{(k)}$ is the attenuation factor between the BS and the user $k$ along path $l$, and $\gamma^{(k)}$ and $\tilde{\gamma}^{(k)}$ are power gains. $L^{(k)}$ are the number of propagation paths between user $k$ and the BS.
\added[id=msb]{In this work, we will choose $\gamma^{(k)} = 1/\alpha_l^{(k)}$, to compensate for the medium  attenuation, and $\tilde{\gamma}^{(k)}$ is a power gain to be applied at the transmitter or BS, and that can be identified as being the linear coefficients of the superposition of chaotic signals.}

I will now consider the uplink, \added[id=msb]{where 2 users send signals that are linearly composed by a superposition that happens at the BS}, each user \added[id=msb]{or source signal is identified with an index} $k=\{1,2\}$, and will in most of the following results neglect in Eq. (\ref{multipath-returnmap-MSB}) the contribution from other propagation paths other than the direct ($L^{(1)}=L^{(2)}=1$).  
Assume user 1 to operate at frequency $f^{(1)}=f=1/T$ and user 2 at frequency $f^{(2)}=2f=2/T$, and $\gamma^{(k)}=1$. In order to reduce the continuous mathematical description of the uplink communication, including the decoding phase to the 2D unfolded Baker's map, I will only treat cases for which \added[id=msb]{the natural frequency of user $k$ is given by} $f^{(k)} = 2^{m} f$, with $m \in \mathbb{N}$, the parameter $\beta^{(k)}=f^{(k)}\ln{(2)}$, \added[id=msb]{and $f$ is the base frequency of user 1, which will be chosen to be 1}. At time $(n+1)T$, the signal received by BS from user $k$=1 as a function of the signal received at $nT$ is described by 
\begin{equation}
r_{n+1}^{(1)} = 2 r^{(1)}_n - \alpha_0^{(1)}s^{(1)}_n. 
\label{user1}
\end{equation}
\noindent

At time $(n+1)T$, the signal received by BS from user $k$=2 as a function of the signal received at 
$nT$ is 
\begin{equation}
r_{2n+2}^{(2)} = 4 r^{(2)}_{2n} - \alpha_0^{(2)}[2s^{(2)}_{2n} + s^{(2)}_{2n+1}], 
\label{user2}
\end{equation}
\noindent
where the $r_{2n}$ represents the value of $r^{(2)}(t=nT)$ (recall that at each time-interval $T$, user 2 chaotic system completes two full cycles each with period $T/2$). Notice that the LE of Eq. (\ref{user2}) will provide a quantity in term of 2 cycles of user 2, \added[id=m]{but 1 cycle in terms of user 1}. So, the LE of  Eq. (\ref{user2})  is equal to $\ln{(4)}$ nepits per each period $T$, which is twice the LE of Eq. (\ref{user1}) for that same period $T$. \added[id=m]{Comparison of both LEs become easier if we calculate them in units of nepits per unit of time. LE for user 1 is $\beta^{(1)}=f^{(1)}\ln{(2)}=\ln{(2)}$ and that for user 2 is $\beta^{(2)}=f^{(2)}\ln{(2)}=2\ln{(2)}$.} This is because user 2 has a frequency twice larger than that of user 1 \cite{eckmann1985ergodic}. Since these two maps are full shift, their LE equals their Shannon Entropy, so their LE represents the encoding capacity (in units of nepit).
Doing the coordinate transformation $r^{(1)}_{n}=2u_n^{(1)}-1$ (for the map in (\ref{user1})) and 
$r^{(2)}_{2n}=2u_n^{(2)}-1$ (for the map in (\ref{user2})) and choosing $\gamma^{(k)}=1/\alpha^{(k)}$, Eqs. (\ref{user1})
and (\ref{user2}) become respectively 
\begin{eqnarray}
u^{(1)}_{n+1}=2u^{(1)}_n-\lfloor 2u^{(1)}_n \rfloor \equiv 2u^{(1)}_n- b^{(1)}_n \label{user1-shift} \\
u^{(2)}_{n+1}=4u^{(2)}_n-\lfloor 4u^{(2)}_n \rfloor \equiv 4u^{(2)}_n- b^{(2)}_n, \label{user2-shift}
\end{eqnarray}
\noindent
where $u^{(k)}_n \in [0,1]$ (in contrast to $r_n^{(k)} \in [-1,1]$), and $b^{(1)}_n=1/2(s^{(1)}_n+1) \in (0,1)$, 
and $b^{(2)}_n=(s^{(2)}_n+s^{(2)}_{n+1}/2) \in (0,1,2,3)$. Equation (\ref{user1-shift}) is simply the Bernoulli shift map, representing the discrete dynamics of user 1 (the signal received after equalizing for the attenuation), and Eq. (\ref{user2-shift}) is the second iteration of the shift map 
representing the discrete dynamics of user 2, (after equalizing the attenuation, \added[id=msb]{by doing $\gamma^{(k)}=1/\alpha^{(k)}$}). 

\begin{figure}[hbt]
\centering
{\includegraphics[height=3.5cm,width=8cm]{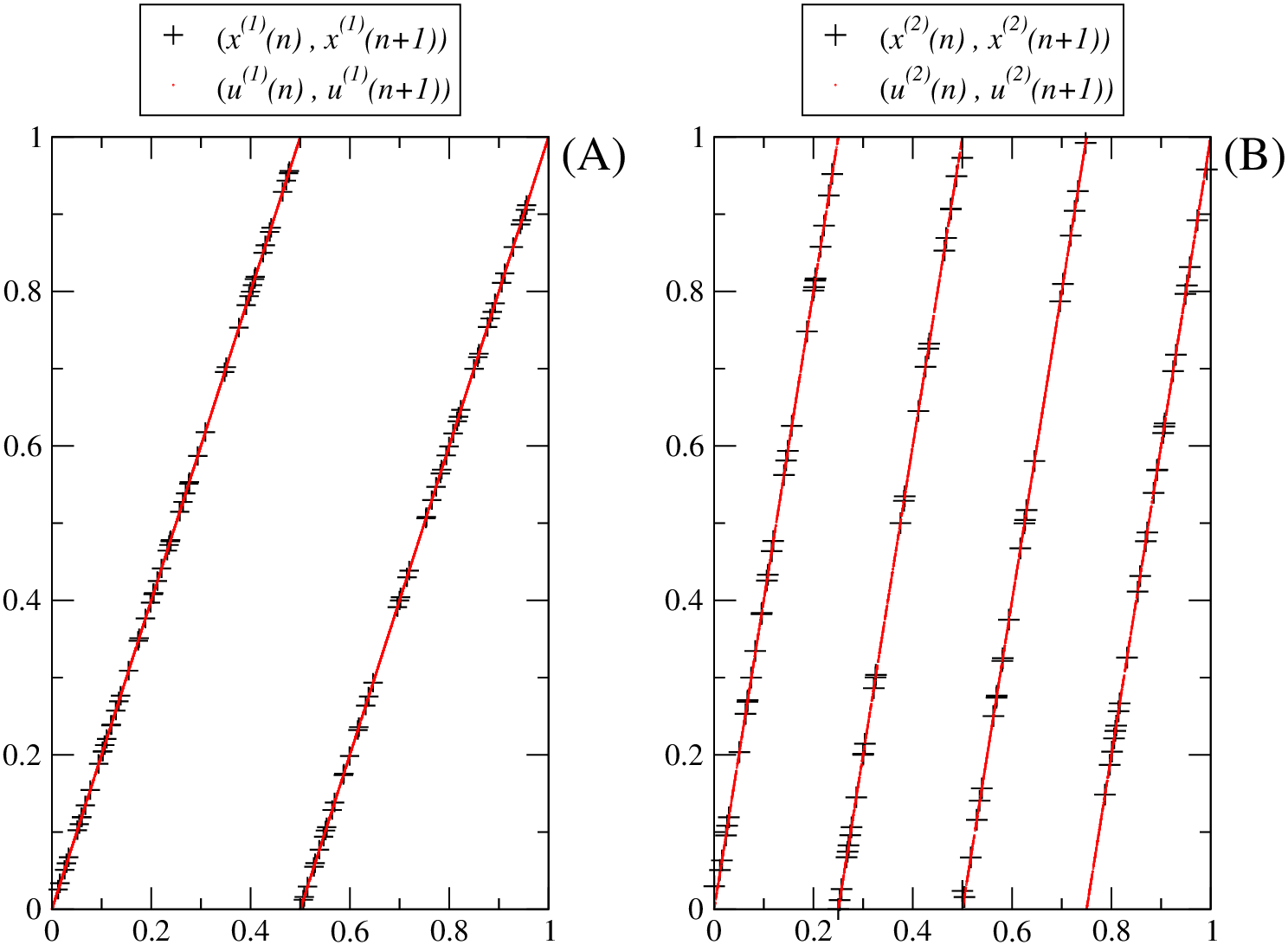}}
\caption{[Colour online] The return maps of Eqs. (\ref{user1-shift}) and (\ref{user2-shift}) are shown by red dots, and corresponding return maps of discrete sets obtained directly from the continuous solution of Eq. (\ref{rescaled-corron}) are shown by black crosses. In (A), discrete states for user $k=1$, and in (B), discrete states for user $k$=2.}
\label{paper-fig1a-b}
\end{figure}

\added[id=msb]{
Figure \ref{paper-fig1a-b}(A)-(B) shows in red dots solutions for Eqs. (\ref{user1-shift}) and (\ref{user2-shift}), respectively. Corresponding return maps of the discrete set of points $x_n^{(k)}$ constructed directly from the continuous solution of Eq. (\ref{rescaled-corron}) with frequency given by 
$f^{(k)}=kf$ by taking points at the time $t=nT$, and doing the normalization as before $x^{(k)}_{n}=2x_n^{(k)}-1$ (so, 
$x_n^{(k)} \in [0,1]$) is shown by the black crosses. }

The composed received signal at \added[id=msb]{discrete times} $nT$, \added[id=msb]{a linear superposition of 2 chaotic signals with different power spectrum,} is given by 
\begin{equation}
O_{n}=\tilde{\gamma}^{(1)}u^{(1)}_n +   \tilde{\gamma}^{(2)}u^{(2)}_{n}. 
\label{combined_received}
\end{equation}
\noindent
\added[id=msb]{Generalization for $N$ source signals can be written as $O_{n}=\sum_{k=1}^{N} \tilde{\gamma}^{(k)}u^{(k)}_n$.} 
At the BS, the received signal is $O_n + w_n$, so it is corrupted by an AGWN $w_n$ that has a signal to noise rate (SNR) in dB as compared with the power of the signal $O_n$. The received \added[id=msb]{discrete-time} return map, for $w_n=0$, \added[id=msb]{can be derived by putting Eqs. (\ref{user1-shift}) and (\ref{user2-shift}) into Eq. (\ref{combined_received})}
\begin{eqnarray}
O_{n+1} &=& 4 O_n - 2 \tilde{\gamma}^{(1)}u^{(1)}_n - \tilde{\gamma}^{(2)}b_n^{(2)} - \tilde{\gamma}^{(1)}b_n^{(1)}
\label{combined_received-return}\\
u^{(1)}_{n+1}&=&2u^{(1)}_n- b^{(1)}_n \label{user1-shift-copy}, 
\end{eqnarray}
\noindent where Eq. (\ref{user1-shift-copy}) is just Eq. (\ref{user1-shift}). 

\subsection{Preservation of LEs for linear compositions of chaotic source signals}\label{preservation-LEs}

The system of Eqs. (\ref{combined_received-return}) and (\ref{user1-shift-copy}) has two distinct positive LEs, one along the direction ${\bm{v}}^{(1)} = (0\mbox{\ }1)$ associated with the user 1 and equal to $\chi^{(1)}=\ln{(2)}$ nepit per period $T$, and another along the direction 
${\bm{v}}^{(2)} = (1\mbox{\ } 0)$, which can be associated with the user 2 and equals $\chi^{(2)}=\ln{(4)}=2ln{(2)}$ nepit per period $T$. 

To calculate the LEs of this \added[id=msb]{2-dimensional} system (see \cite{maria2019}) we consider the expansion of a unitary basis of orthogonal perturbation vectors $\bf{v}$ and calculate them by  
\begin{equation}
    {\bm{\chi}} = \lim_{n\to\infty} \frac{1}{n} \ln{\norm{  \bm{M} \cdot \bf{v} }}
    \label{LE-combined-received}, 
\end{equation}
where $\norm{\bm{v}}$ is the norm of vector $\bf{v}$, $\bm{M}=\bm{J}^n$, and $\bm{J}=\left(\begin{array}{cc} 4 & -2\tilde{\gamma}^{(1)} \\
0 & 2\end{array} \right)$. Thus, combining chaotic signals with different frequencies as \added[id=msb]{a linear superposition described by } Eq. (\ref{combined_received}) preserves the spectra of LEs of the signals from the users alone. \added[id=msb]{This is a hyperbolic map where the sum of the positive Lyapunov exponents is equal to the Kolmogorov-Sinai's entropy, which represents the information rate.}  Consequently, the information received is equal to the sum of the information transmitted by both users, for the no noise scenario. \added[id=m]{More details about this relationship is presented in the next Sec. \ref{LE-and-information}}. \added[id=msb]{In other words, a linear superposition of chaotic signals as represented by Eq. (\ref{combined_received}) does not destroy the information content of each source signal}. Preservation of the spectrum of the LEs in a signal that is a \added[id=msb]{linear superposition of chaotic signals with different power spectrum} is a universal property of chaos. Demonstration is provided in \textbf{Appendix C}, \added[id=msb]{where I study signals composed by two variables from the R\"ossler attractor, user 2 with a base frequency that is $Q$ times that of the user 1.  This demonstration uses an equivalence principle.}  
Every wireless communication network with several users 
can be made equivalent to a single user in the presence of several 
imaginary propagating paths. Attenuation and power gain factors need to be recalculated to compensate for a signal that is in reality departing from user 2 but that is being effectively described as departing from user 1.
Suppose the 2 users case, both with the same frequency $f^{(k)}=f$, in the uplink scenario. The trajectory of user 2 at a given time $t$, $x^{(2)}(t)$, can be described in terms of the trajectory of the user 1 at a given time $t-\tau$. 
\added[id=msb]{So, the linear superposition of 2 source signals in Eq. (\ref{WiChaos_up}) can be simply written as a single source with time-delayed components as}
\begin{eqnarray}
O(t)_{u}    = \sum_{l=0}^{L^{(k)}-1} [\alpha_l^{(1)} \gamma^{(1)}  \tilde{\gamma}^{(1)} x^{(1)}(t-\tau_l^{(1)}) \nonumber \\
+ \alpha_l^{(2)} \gamma^{(2)}  \tilde{\gamma}^{(2)} x^{(1)}(t-\tau_l^{(1)}-\tau)]  
+ w(t). 
\label{WiChaos_up-equiv}
\end{eqnarray}

\added[id=msb]{In practice, $\tau$ can be very small, because of the sensibility to the initial conditions and transitivity of chaos. For a small $\tau$ and $\epsilon$ 
it is true that
$|x^{(2)}(t)  - x^{(1)}(t-\tau)| \leq \epsilon$, regardless of $t$.}

This property of chaos is extremely valuable, since when extending the ideas of this work to arbitrarily large and complex communicating networks, one might want to derive expressions such as in Eqs. 
(\ref{combined_received-return}) and (\ref{user1-shift-copy}) to decode the information arriving at the BS.
Details of how to use this principle to derive these equations 
for two users with $f^{(2)}=2f^{(1)}$, and also when $f^{(2)}=f^{(1)}$ are shown in 
\textbf{Sec. II of SM}. 

\subsection{Lyapunov exponents, the information carried by chaotic signals and the information capacity of Wi-C1}\label{LE-and-information}

\added[id=m]{

Pesin's equality relates positive Lyapunov exponents (LEs) with information rate of a chaotic trajectory \cite{pesin2008dimension}: The sum of positive LEs of a chaotic trajectory is equal to the Kolmogorov-Sinan entropy, denoted $H_{KS}$ (a kind of Shannon entropy rate), a quantity that is considered to be the physical entropy of a chaotic system. 
This is always true for chaotic systems that possess the Sinai-Ruelle-Bowen (SRB) measure \cite{young2002srb}, or more precisely that have absolutely continuous conditional measures on unstable manifolds. In this work, I have considered a parameter configuration such that the system used to generate chaotic signals is described by the shift map, a hyperbolic map, which has SRB measure. Therefore, the amount of information transmitted by a user is given by the LE of the system in Eq. (\ref{rescaled-corron}). 

I have demonstrated that linearly composed chaotic signals with different natural frequencies preserve all the positive LEs of the source signals (Appendix C). By a chaotic signal I mean a 1-dimensional scalar time-series, or simply a single variable component of a higher dimensional chaotic trajectory. If the chaotic signals are generated by  Eq. (\ref{rescaled-corron}), their linear composition in Eqs. (\ref{combined_received-return}) and (\ref{user1-shift-copy}) is still described by a hyperbolic dynamics (possessing SRB measure), and thus leading to a trajectory whose information content is given by the sum of the positive LEs, which happens to be equal to the sum of the LEs of the source signals.  So,  the information encoding capacity in units of nepits per unit of time of the Wi-C1, denoted by $\mathcal{C}_e$,  when users use the system in Eq. (\ref{rescaled-corron}) to generate chaotic signals is given by the sum of Lyapunov exponents of the source signals:}

\begin{equation}
\mathcal{C}_e = \sum_k f^{(k)} \beta(\gamma=1)^{(k)} = \sum_k \beta(\gamma)^{(k)}
\label{information-capacity}
\end{equation}
\noindent
\added[id=m]{
where $ f^{(k)}$ and $\beta(\gamma=1)^{(k)}$, and are the natural frequency of the signal and the LE of user $k$ (in units of nepits per unit of time), respectively. By information encoding capacity, I mean the information rate of a signal that is obtained by a linear composition of chaotic signals. If linear coefficients (power gains) are appropriately chosen (see next Sec. \ref{channel-encoding}) and noise is sufficiently low (see Sec. \ref{performance}), then the information encoding capacity of Wi-C1 is equal to the information capacity of Wi-C1, or the total rate of information being received/decoded.}

\added[id=m]{

It is worth discussing however what would be the information capacity of Wi-C1, in case one considers users communicating with other chaotic systems 
than that described by Eq. (\ref{rescaled-corron}). My result in Appendix C demonstrates that all the positive Lyapunov exponents of the chaotic source signals 
are present in the spectra of the linearly composed chaotic signals constructed using different chaotic signals (that may have different natural frequencies), and 
being generated by the same chaotic system. 

Recent work \cite{catalan2019,matsuoka2015} has shown that there is a strong link between the sum of the positive LEs and the topological entropy, denoted $H_T$, in a chaotic system. The topological entropy measures the rate of exponential growth of the number of distinct orbits, as we consider orbits with growing periods. For Eq. 
(\ref{rescaled-corron}), its topological entropy equals its positive LE and its Kolmogorov-Sinai entropy. So, $H_T = \beta(\gamma)=H_{KS}$ (in units of bits per unit of time). That is not always the case. Denoting the sum of LEs of a chaotic system by $\sum^+ $, one would typically expect that  $H_T  \geq H_{KS}$ and moreover that $\sum^+  \geq H_{KS}$. However, the recent works in  Refs.  \cite{catalan2019,matsuoka2015} have shown that there are chaotic systems for which $H_T  = \sum^+ $.

This work considers that the proposed communication system Wi-C1 has users that use chaotic signals generated by means of controlling (class (i) discussed in Sec. \ref{Introduction}), so that the trajectory can represent the desired information to be transmitted. 
The work in Ref. \cite{hayesPRL1993} has shown that the information encoding capacity of a chaotic trajectory produced by control 
is given by the topological entropy of the non-perturbed system, not by its Kolmogorov-Sinai entropy. Therefore, if only a single user is being considered in the communication (e.g., only one transmitter), and this user generates chaotic signals for which $H_T  = \sum^+ $, the information encoding capacity of this communication system would be given by $\sum^+$.

Let us now discuss the multi-user scenario, still assuming that the users generate their source chaotic signals using systems for which $H_T = \sum^+ $.  As demonstrated in Appendix C, all the positive Lyapunov exponents of chaotic source signals are preserved in a linearly composed signal. Moreover, since that LEs of a chaotic signal are preserved by linear transformations, and since a linear transformation to a signal does not alter its information content, it is suggestive to consider that the information capacity of this multi-user communication system would be given by the sum of the positive LEs of the chaotic source signals for each user. This however will require further analysis.}

\subsection{Preparing the signal to be transmitted (encoding): finding appropriate power-gains}\label{channel-encoding}

In order to avoid interference \added[id=msb]{or false near neighbours crossing in the received composed signal}, allowing one to discover the symbols $b^{(1)}$ and $b^{(2)}$ only by 
observing the 2-dimensional return map of $O_{n+1} \times O_{n}$ \added[id=msb]{that maximises the separation among the branches of the map to avoid mistakes induced by noise, we need to appropriately choose the power gains}
$\tilde{\gamma}^{(k)}$.
Looking at the mapping in Eq. (\ref{combined_received-return}), the term $2^{f^{(2)}} O_n$ represents a piecewise linear map with $2^{f^{(2)}}$ branches. The spatial domain for each piece has a length denoted by $\zeta(f^{(2)})$. The 
term $(2^{f^{(2)}} - 2) \tilde{\gamma}^{(1)} u^{(1)}_{n}$ representing the dynamics for the smallest oscillatory frequency is described by a piecewise linear map with $(2^{f^{(2)}} - 2)$ branches. To avoid interference, the return map for this term must occupy a length $\zeta(f^{(1)})$ that is fully embedded within the domain for the dynamics representing higher order frequencies. Assuming that for a given number of users $N$, all frequencies $f^{(i)}$ with $i=1,\ldots,N$ are used, this idea can be expressed in terms of an equation where
\begin{equation}
\zeta(f^{(i)}) = 2^{(f^{(i)})} \zeta(i-1), \mbox{\ \ } i=\{1, ..., N\}.
\label{scales}
\end{equation}
Then, $\tilde{\gamma}^{(k)}=\zeta(k)$, but for a received map within the interval $[0,1]$,  normalization of the values o   $\tilde{\gamma}^{(k)}$ by 
\begin{equation}
\tilde{\gamma}^{(k)} = \frac{\zeta(k)}{\sum_{i=1}^N \zeta(f^{(i)})}. 
\label{power-gains}
\end{equation}
For \added[id=msb]{2 users} ($N=2$) and $\zeta(1)=0.2$, the \added[id=msb]{appropriate power gains to be chosen in the encoding phase and that allows for the decomposition (or decoding) of the information content of the composed received signal are given by} $\tilde{\gamma}^{(1)}=0.2$ and $\tilde{\gamma}^{(2)}=0.8$. Using these values for $\tilde{\gamma}^{(1)}$ and $\tilde{\gamma}^{(1)}$ in Eq. (\ref{combined_received}) and considering an AWGN $w_n$ with SNR of 40dB (with respect to the power of $O_n$) produces the return map shown by points in Fig. \ref{paper-figure1c-e}(A), with 8 branches \added[id=msb]{all aligned along the same direction (the branches would have the same derivative for the no noise scenario), which therefore prevents crossings or false near neighbours - and are also equally separated to avoid mistakes in the decoding of the information due to noise.}

The choice of the power gains for the {\bf{downlink configuration}} is similarly done as in the uplink configuration, taking into consideration that  each user has its own noise level. This is shown in \textbf{Sec. III of SM}.

\section{Decomposing the linear superposition of chaotic signals, and the decoding of signals and their information content.}\label{decoding}

\subsection{Decomposition (decoding) by thresholding received signal}


Communication based on chaos offers several alternatives \added[id=m]{for decoding, or in other words, the process to obtain the information that is conveyed by the received signal}. 
Assuming the received signal is modelled by  Eqs. (\ref{combined_received-return}) and (\ref{user1-shift-copy}), with the appropriated power gains as in Eq. (\ref{power-gains}),  the optimal 2-dimensional partition to decode the digital information is 
described by the same map of Eqs. (\ref{combined_received-return}) and (\ref{user1-shift-copy}) with a 
translation.  For the case of 2 users in the uplink scenario, this translates into a 7 lines partition 
\begin{eqnarray}
O^*_{n+1}(j)&=&4O^*_n(j) - T_j, \label{decoding-partition} \\
T_j&=&\frac{1}{2}\left[ 3 \tilde{\gamma}^{(1)} + (j-1)\tilde{\gamma}^{(2)} \right], \mbox{\ \ } j=\{1,\ldots,7\}. \nonumber 
\end{eqnarray}
These partition lines for $\tilde{\gamma}^{(1)}=0.2$ and $\tilde{\gamma}^{(2)}=0.8$ are shown by the coloured straight lines in Fig. \ref{paper-figure1c-e}(A). \added[id=msb]{They allow for the decomposition/decoding of the digital (symbolic) information contained in the composed received signal.}

\subsection{Decomposition (decoding)  by filtering received signal}

A more sophisticated approach to decode information is based on a matched filter \cite{corron2010matched}. 
\added[id=msb]{In here I show that the system formed by Eq. (\ref{rescaled-corron}) and its matched filter can be approximately described by the  unfolded Baker's map, a result that allows us to understand that the recovery of the signal sent by a user from the composed signal solely depends on the inverse dynamics of this user.}
Details of the fundamentals presented in the following can be seem in \textbf{Sec. IV of SM}.    
If the equations describing the dynamics of the transmitted chaotic signal (in this case Eq. (\ref{rescaled-corron})) possess no negative Lyapunov exponents  - as it is shown \textbf{Sec. I of SM} - attractor estimation of a noisily corrupted signal can be done using its time-inverse dynamics that is stable and possess no positive LEs (\textbf{shown in Sec. V of SM}). The 
evolution to the 
future of the time-inverse dynamics 
is described by a system of ODE hybrid equations obtained by the time-rescaling $d/dt^{\prime} = -d/dt$ applied to 
Eq. (\ref{rescaled-corron}) resulting in  
\begin{equation}
    \ddot{y}+2\beta\dot{y}+(\omega^2+\beta^2)[y-\eta(t)]=0, 
\label{rescaled-corron-inverse}
\end{equation}
\noindent
\noindent
where the variable $y$ represents the $x$ in time-reverse, and
as shown in \textbf{Sec. IV of SM},  if 
$\eta(t)$ is defined by 
$\dot{\eta(t)}=x(t)-x(t-T)$ (defined as $\dot{\eta(t)}=x(t+T)-x(t)$ in Ref. \cite{corron2010matched}) it can be roughly approximated to  be equal to the symbol $s(t)$. 

Taking the values of $y$ at discrete times at $nT$, writing that $y(nT)=y_n$, and defining the new variable for users 1 and 2 as before  $y^{(1)}_n=2z^{(1)}_n -1$ and $y^{(2)}_{2n}=2z^{(2)}_n -1$
if Eqs. (\ref{user1-shift}) and (\ref{user2-shift}) are map solutions of Eq. (\ref{rescaled-corron}) (in the re-scaled coordinate system, with appropriate $\gamma$ gains) for user $k$ with frequencies 
$f^{(k)}=k$, their inverse mapping the solution of Eq. (\ref{rescaled-corron-inverse}) is given by 

\begin{equation}
z^{(k)}_{n+1}=2^{-k}\{z^{(k)}_n-\lfloor 2^{k}u^{(k)}_n \rfloor\}, \mbox{\ and } \lfloor 2^{k}u^{(k)}_n \rfloor \equiv b^{(k)}_n,  
\label{inverse-map}
\end{equation}
This map can be derived simply defining $z^{(k)}_{n+1} = u^{(k)}_{n}$ and  
$z^{(k)}_{n} = u^{(k)}_{n+1}$. We always have that $\lfloor 2^{k}u^{(k)}_n  \rfloor = b^{(k)}_n$. So, for any 
$z^{(k)}_n \in [0,1]$ and which can be simply chosen to be equal to the received composed signal $O_n$ (normalized such that $\in [0,1]$), it is also true that  
\begin{equation}
\lfloor 2^{k}    z^{(k)}_{n+1} \rfloor = \lfloor 2^{k}u^{(k)}_n  \rfloor = b^{(k)}_n.  
\label{decoding-by-variable}
\end{equation}
\noindent
So, if we represent an estimation of the transmitted symbol of user $k$ by $\tilde{b}^{(k)}_n$, then decoding of the transmitted symbol of user $k$ can be done by calculating $z^{(k)}_{n+1}$ using the inverse dynamics of the user $k$
\begin{equation}
z^{(k)}_{n+1}=2^{-k}\{z^{(k)}_n - \tilde{b}^{(k)}_n\}.   
\label{inverse-map-decode3}
\end{equation}
\noindent
and applying this value to Eq. (\ref{decoding-by-variable}).  
This means that the system formed by the variables 
$u^{(k)}_{n},z^{(k)}_{n}$ is a generalization (for $k \neq 1$)  of the unfolded Baker's map \cite{lasota1985probabilistic}, 
being described by a time-forward 
variable $u^{(k)}_{n}$ (the Bernoulli shift for $k$=1), and its backward  variable component $z^{(k)}_{n}$. 

\begin{figure}[hbt]
\centering
{\includegraphics[scale=0.32]{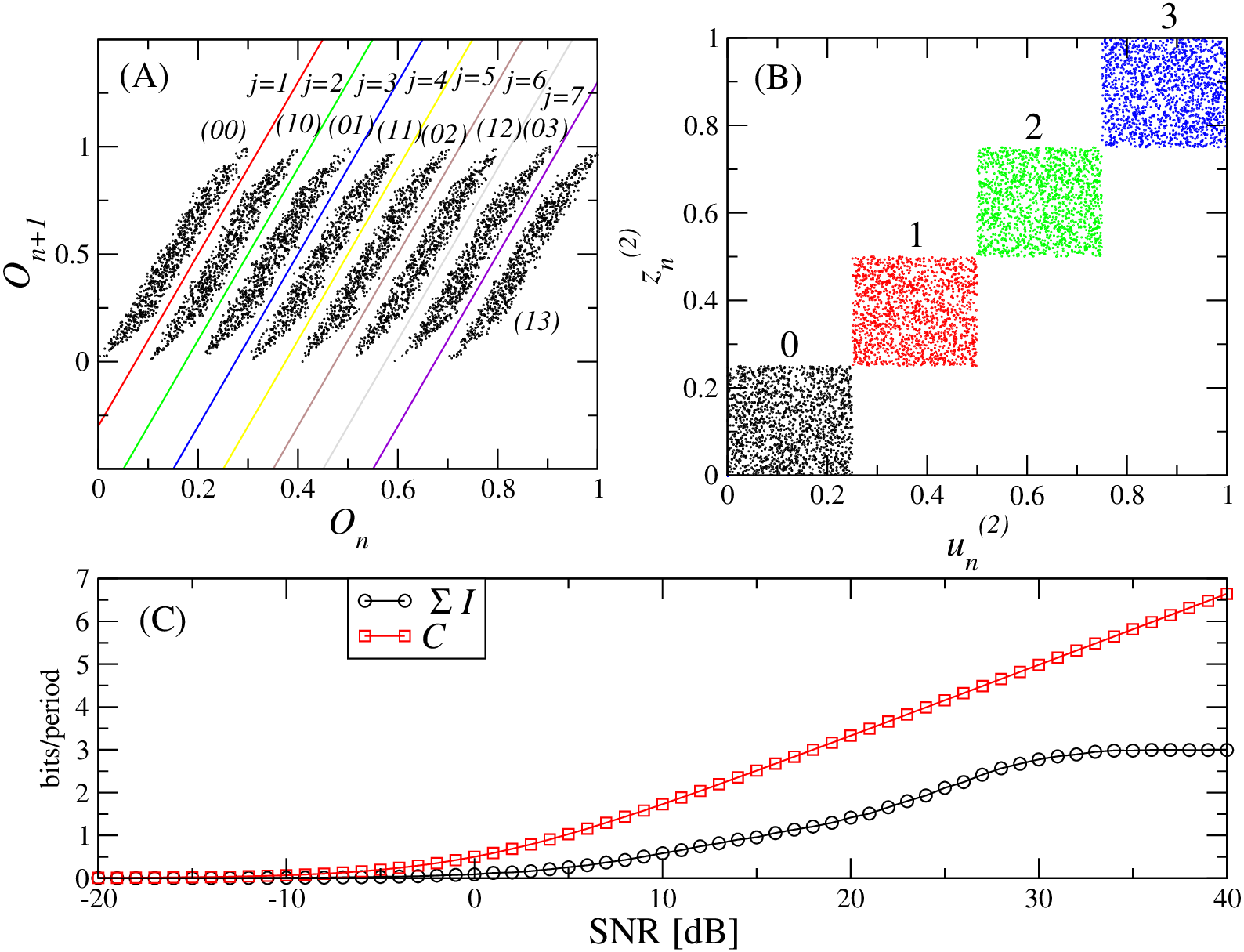}}
\caption{[Colour online] (A) points shows the return map of the received signal with $\tilde{\gamma}^{(1)}=0.2$ and $\tilde{\gamma}^{(2)}=0.8$, and the lines the partitions from which received symbols are estimated. Inside the parenthesis, the first symbol is from user 2 and the second symbol is from user 2. In (B) one sees a solution of the unfolded Baker's map, where horizontal axis shows trajectory points from Eq. (\ref{user2-shift}) and vertical axis trajectory points from Eq. (\ref{decoding-by-variable}), for the user $k$=2. In (C) is shown $C$ against $\sum I$, with respect to the Signal to Noise Ratio (SNR).}
\label{paper-figure1c-e}
\end{figure}

Figure   \ref{paper-figure1c-e}(B) demonstrates that it is possible to extract the signal of a user (user $k$=2) from the composed signal, $O_n$ (Eq. (\ref{combined_received})), by setting in Eq. (\ref{inverse-map-decode3}) that 
$z^{(2)}_n=O_n$, and $\tilde{b}^{(k)}_n = b^{(k)}_n$. Even thought $u^{(2)}_n \neq z^{(2)}_n$, decoding Eq. (\ref{decoding-by-variable}) is satisfied. 
\added[id=msb]{Therefore, the matched filter that decomposes the source signal of a user from the received composed signal is the matched filter of that user alone.}

\section{Analysis of performance of Wi-C1, under noise constraints}\label{performance}

I can now do an analysis of the performance of the Wi-C1, for both the uplink and the downlink configurations, for 2 users modelled by 
Eqs. (\ref{user1-shift}) and (\ref{user2-shift}) with power gains
$\tilde{\gamma}^{(1)}=0.2$ and $\tilde{\gamma}^{(2)}=0.8$.
The information capacity for both users (in bits per iteration, \added[id=m]{or bits per period}) is given by
\begin{eqnarray}
C&=&0.5\log_2{\left( 1 + SNR \right)},   
\end{eqnarray}
\noindent
where  \added[id=m]{$SNR = \frac{P}{P^{w}}$ (units in dB, decibel)} is the signal-to-noise ratio, the ratio between the power $P$ of the linearly composed signal 
$\tilde{\gamma}^{(1)}u_n^{(1)} + \tilde{\gamma}^{(2)}u_n^{(2)}$ (arriving at the BS, in the uplink configuration, or departing from it, in the downlink configuration) and 
$P^{w}$, the power of the noise $w_n$ at the BS (for the uplink configuration) or at the users (for the downlink configuration, assumed to be the same).
The total capacity of the communication denoted by $C$ is calculated assuming that decoding of users 1 and 2 are simultaneously done from the noisily corrupted received signal $O_n+w_n$ (see Eq. (\ref{combined_received})), and so, decoding of the signal from user 2 does not treat the signal of user 1 as noise. 

This capacity has to be compared to the actual 
rate of information being realised at the BS (or at the receivers), quantified by the Mutual Information, $I(b_n^{(k)};\tilde{b}_n^{(k)})$ between the symbols transmitted ($b_n^{(k)}$) and the decoded symbols $\tilde{b}_n^{(k)}$ estimated by using partition in Eq. (\ref{decoding-partition}), defined as usual by 
$I(b_n^{(k)};\tilde{b}_n^{(k)}) = 
H(b_n^{(k)}) - H(b_n^{(k)}|\tilde{b}_n^{(k)}) $
\noindent
where $H(b_n^{(k)})$ denotes the Shannon's Entropy of the user $k$ which is equal to the positive LE of the user $k$, for $\beta(\gamma=1) = \ln{(2)}$, and   $H(b_n^{(k)}|\tilde{b}_n^{(k)})$ is the conditional Entropy.  

Figure \ref{paper-figure1c-e}(C) shows in red squares the full theoretical capacity given by $C$ against the rate of information decoded given by  $\sum I = I(b_n^{(1)};\tilde{b}_n^{(1)}) + I(b_n^{(2)};\tilde{b}_n^{(2)})$, in black circles, with respect to the SNR. As it is to be expected, the information rate received $\sum I$ is equal to \added[id=m]{the information encoding capacity $\mathcal{C}_e$ that is transmitted} (both equal to 3bits/period) for low noise levels, tough smaller than the theoretical limit.

Notice that this analysis was carried out using the map version of the matched filter \cite{corron2010matched} in Eq. (\ref{rescaled-corron-inverse}), and as such lacks the powerful use of the negativeness of the LE to filter noise. Moreover, decoding used the trivial 2D threshold by Eq. (\ref{decoding-partition}), and not higher-dimensions reconstructions. 

\subsection{Comparison of performance of Wi-C1 against NOMA}\label{Wi-C1-versus-Noma}

\begin{figure}[hbt]
\centering
{\includegraphics[scale=0.32]{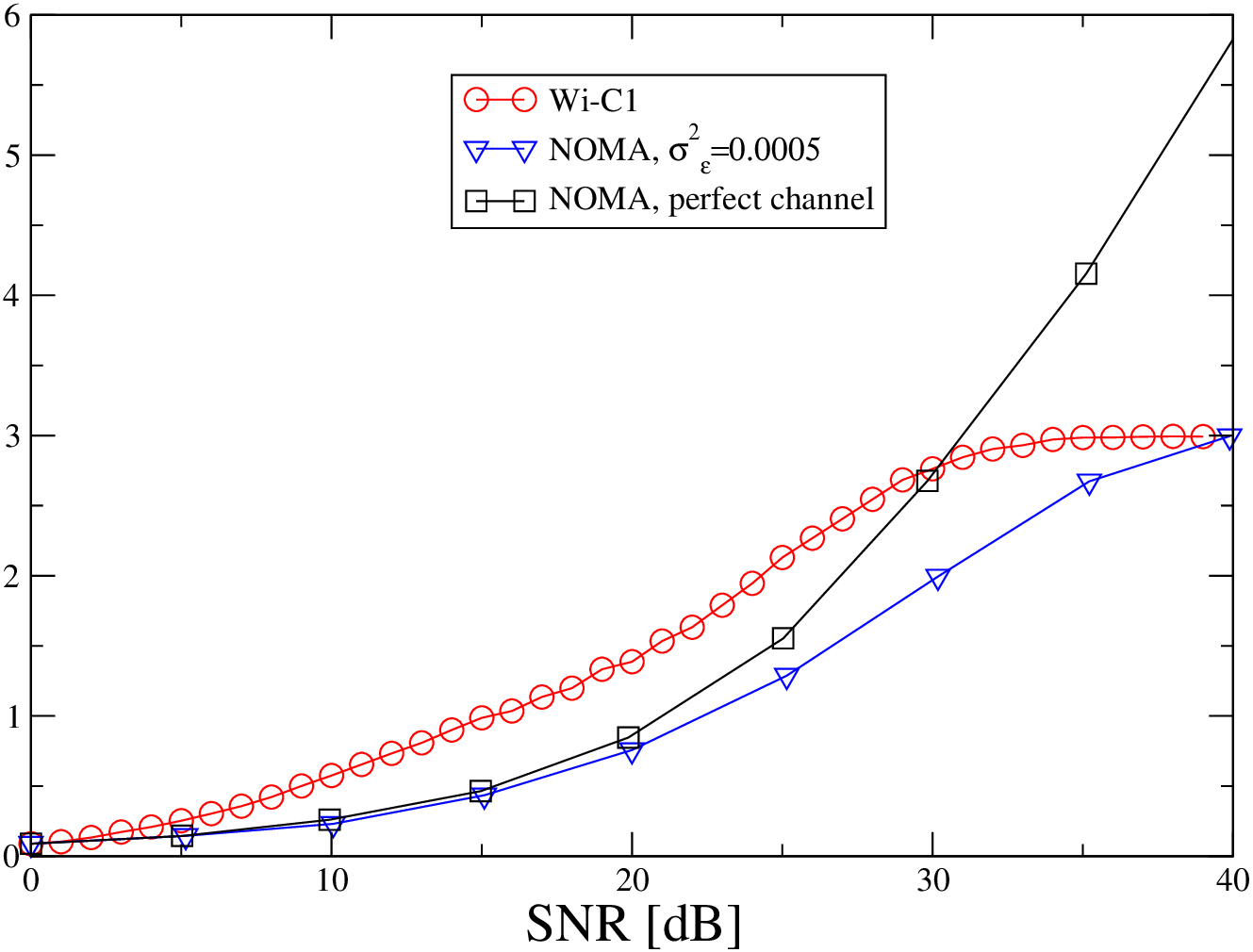}}
\caption{[Colour online] Red circles show $\sum I$, blue down triangle show the average sum rate for partial channel knowledge ($\sigma_{\epsilon}=0.0005$), and  black squares show the average sum rate for perfect channel knowledge, with respect to the Signal to Noise Ratio (SNR).}
\label{paper-fig3}
\end{figure}

\added[id=m]{To cope with the expected demand in 5G wireless communication, non-orthogonal multiple access (NOMA) \cite{kizilirmak2016non,benjebbour2017overview,dai2018survey} was proposed to allow all users to use the whole available frequency spectrum. One of the most popular NOMA scheme allocate different power gains to the signal of each user. Full description of this scheme and its similarities with Wi-C1 is given in Sec. VI of SM. 

The key concept behind NOMA is that users signals are superimposed 
with different power gains, and successive
interference cancellation (SIC) is applied at the user with better
channel condition, in order to remove the other users signals
before detecting its own signal \cite{Saito2013}. In the Wi-C1, as well as in NOMA, power gains are also applied to construct the linear superposition of signals. But in this work, I assume that the largest power gain is applied to the user with largest frequency. Moreover, in this work, I have not done successive interference cancellation (SIC), since the information from all the users are simultaneously recovered by the thresholding technique, by considering a trivial 2D threshold by Eq. (\ref{decoding-partition}). 

Comparison of the performance of Wi-C1 and  NOMA is done considering the work in Ref. \cite{YangIEEE2016}, which has analysed  
the performance of NOMA for 2 users in the downlink configuration, under partial channel knowledge. Partial channel knowledge means in rough terms that the "amplitude" of the signal arriving to a user from the BS is incorrectly estimated. In this sense, I have considered in the Wi-C1 perfect channel knowledge, since 
my simulations in Fig. \ref{paper-figure1c-e}(C) based on Eq. (\ref{combined_received}) assumes that $\gamma^{(k)}=\frac{1}{\alpha^{(k)}}$ to compensate for the amplitude decay $\alpha^{(k)}$ in the physical media (see Eq. (\ref{WiChaos_up})). More precisely, partial channel knowledge means that a Gaussian distribution describing the signal amplitudes departing from a user decreases its variance inversesly proportional to a power-law function of the distance between that user and BS. The variance of the error of this distribution estimation is denoted by $\sigma_{\epsilon}$, an important parameter to understand the results in Ref. \cite{YangIEEE2016}. Partial channel knowledge will impact on the optimal SIC performed for the results in Ref. \cite{YangIEEE2016}. Recall again that for the Wi-C1, no SIC is performed. 

In Fig. \ref{paper-fig3}, the curve for $\sum I$ (the rate of decoded information) in Fig. \ref{paper-figure1c-e}(C) is plot in red circles and compared with data shown in Fig. 3 of Ref. \cite{YangIEEE2016} for the quantity ``average sum rate'', where each dataset considers a different channel configuration. Blue down triangles show the quantity 
``average sum rate" for perfect channel knowledge ($\sigma_{\epsilon}=0$), and black squares represent the same quantity for partial channel knowledge ($\sigma_{\epsilon}=0.0005$). The data points in Fig. 3 of Ref. \cite{YangIEEE2016} were extracted by a digitalisation process. The quantity $\sum I$ for Wi-C1 in in Fig. \ref{paper-figure1c-e}(C) in units of 
bits per period (or channel use) is compared with the quantity ``average sum rate" (whose unit was given in bits per second per Hz) by assuming the period of signals in 
Ref. \cite{YangIEEE2016} is 1s. The average value obtained in  Ref. \cite{YangIEEE2016} has taken into consideration Monte Carlo simulations of several configurations for 2 users that are uniformly distributed in a disk
and the BS is located at the center.}

\added[id=m]{The results in Fig. \ref{paper-fig3} show that Wi-C1 has similar performance in terms of the bit rate for 0dB, better performance for the SNR $\in ]0,30]dB$ as that of the   
NOMA (with respect to the average sum rate) for perfect channel knowledge, and better performance for the SNR $\in ]0,40[dB$ as that of the   
NOMA (with respect to the average sum rate) for partial channel knowledge.

One needs to have into consideration that this outstanding performance of Wi-C1 against NOMA is preliminary, requiring more deep analysis, 
but that is out of the scope of the present work.  
}

\section{Other non-ideal physical media}\label{NL-channels}

\added[id=m]{Previous Sections of this work have tackled with great rigour and detail how chaotic signals are affected when travelling through medium that presents non-ideal  properties such as multipath propagation, noise and chaotic interference (linear superposition), and how this impacts on the proposed communication system. This Section is dedicated to conceptually discuss with some mathematical support how chaotic signals and their information content are transformed by physical channels with other non-ideal properties (dispersion and interference with periodic signals), and how this impacts on the multi-user communication system proposed. 

For the following analysis, I will neglect the existence of multiple indirect paths of propagation, and will consider that only the direct path contributes to the transmission of information, so $L=1$. 
I will consider the uplink scenario where users transmit to a BS. 
I will initially focus the analysis about the impact of the non-ideal physical medium on the signal of a single user, in particular the effect of the medium in the 
received discrete signal being described by Eq. (\ref{multipath-returnmap-MSB}) and  its Lyapunov exponent (LE), and will then briefly discuss the impact of the non-ideal medium on a communication configuration with multi-users.}

\subsection{Physical media with dispersion}

\added[id=m]{Physical media with dispersion are those in which waves have their phase velocity altered as a function of the frequency of the signal. However, a dispersive medium does not alter the frequency of the signal, and therefore it does not alter its natural period, only its propagation velocity.  As a consequence, 
the LE of any arbitrary chaotic signal travelling in a dispersive medium are not modified. The information carried by this chaotic signal would also not be altered, if it were generated by 
Eq. (\ref{rescaled-corron}), or by a system whose chaotic trajectory possesses SRB measure, or that its topological entropy $H_{T}$ equals the sum of the positive LEs. 

However, the travel time of a signal to arrive at the BS along the direct path $\tau_0$ is altered. This can impact on the ability to decode as can be seem from Eq. (\ref{multipath-returnmap-MSB}).  Suppose that the travel time of user $k$, given by $\tau^{(k)}_0$, increases from 0 (as in the previous derivations) to a finite value that is still smaller than the period of that user $T^{(k)}$, so that $n^{\prime}=n$ for that user. But, $\mathcal{K}_0$ would be different than 1, and as a consequence, the return map of the received signal would contain a term that is a function of the symbol $s_{n+1}$. Extracting the symbols from the received discrete signal (decoding) would have to take into consideration this extra symbol, which represents a symbol 1 iteration (or period) in the future. Decoding for the symbol $s_n$ from the received signal would require the knowledge of the symbol $s_{n+1}$. So, to decode what is being received at a given moment in the present would require knowledge of the symbol that has just been sent. To circumvent this limitation, one could firstly send a dummy symbol known by both the transmitter and the receiver at the BS, and use it to decode the incoming symbol $s_n$, which then could be used to decode $s_{n-1}$, and so on. Noise could impact on the decoding. Every new term that appears in Eq.  (\ref{multipath-returnmap-MSB}) results in a new branch for this map. With noise, a branch in the return map that appears due to the symbol $s_{n+1}$ could be misinterpreted  as a branch for the symbol $s_n$, causing errors in the decoding.

In a multi-user scenario, dispersion would only contribute to changing the time-delays $\tau^{(k)}_l$ for each user for each propagating path. As discussed, this will not affect the LEs of the source chaotic signals. Moreover, as demonstrated, the LEs of the source signals should be preserved by the linearly composed signal arriving at the BS, suggesting that the information encoding capacity given by Eq. (\ref{information-capacity}) in the multi-user scenario could also be preserved for the systems for which $H_T=\sum^+$ or $\sum^+ = H_{KS}$ (as discussed in Sec. \ref{LE-and-information}). Noise would however increase the chances of mistakes in the decoding of a multi-user configuration, thus impacting on the information capacity of the communication, since branches in the mapping of the received signal could overlap. At the overlap, one cannot discern which symbol was transmitted.}       

\subsection{Physical media with  interfering period (non-chaotic) signals.}

\added[id=m]{This case could be treated as a chaotic signal that is modulated by a periodic signal. Assuming no amplitude attenuation, the continuous signal arriving at the BS from user $k$ can be described by }
\begin{equation}
r^{(k)}(t)    = x(t) + A\sin{(2\pi f_p t + \phi_0)}
\label{wireless_channel-one-user-modulated}
\end{equation} 
\noindent
\added[id=m]{where $f_p$ represents the frequency of the periodic signal, and $\phi_0$ its initial constant phase. In here, I analyse the simplest case, when $f_p=f^{(k)}$, in which the discrete time signal arriving at the BS at times $t=nT$, from user $k$, would receive a constant contribution $c^{(k)} = A\sin{(2\pi n + \phi_0)}$, due to the interfering periodic signal. If $r^{(k)}_n$ and $r^{(k)}_{n+1}$ denote the discrete time signals arriving at the BS without periodic interference from user $k$ at discrete times $t=nT$ and $t=(n+1)T$, respectively,  then $\tilde{r}^{(k)}_n$ and $\tilde{r}^{(k)}_{n+1}$ described by}
\begin{eqnarray} 
\tilde{r}^{(k)}_n &=& r^{(k)}_n + c^{(k)} \label{tildeRn} \\
\tilde{r}^{(k)}_{n+1} &=& r^{(k)}_{n+1} + c^{(k)} \label{tildeRn1} 
\end{eqnarray}  
\added[id=m]{would represent the discrete time signals arriving at times $t=nT$ and $t=(n+1)T$ at the BS, respectively, after suffering interference from the periodic signal. Substituting these equations into the mapping in Eq. (\ref{multipath-returnmap-MSB}) would allow us to derive a mapping for the signal with interference}
\begin{equation} 
\tilde{r}^{(k)}_{n+1} = e^{\frac{\beta}{f}} \tilde{r}^{(k)}_n - (e^{\frac{\beta}{f}} -1)(c^{(k)} + \alpha_0 s_n). 
\label{return-map-interference} 
\end{equation}  
\added[id=m]{As expected, adding a constant term to a chaotic map does not alter its LE given by $\frac{\beta}{f}$. Consequently, 
the information encoding capacity of this chaotic signal is also not be altered, since it was generated by 
Eq. (\ref{rescaled-corron}). 
 
This constant addition results in a vertical displacement of the map by a constant value  -$(e^{\frac{\beta}{f}} -1) c^{(k)}$. So, added noise in the received signal with interference would not impact more than the impact caused by noise in the signal without interference.  

In a multi-user scenario, LEs of the linearly composed signal arriving at the BS should preserve all the LEs of the source chaotic signals, suggesting that the information encoding capacity in the multi-user scenario could also be preserved, for signals being generated by the chaotic systems discussed in Sec. \ref{LE-and-information}. Noise would however increase the chances of mistakes in the decoding of a multi-user configuration, thus impacting on the information capacity of the communication, since for each user the branches of the mapping describing the received signal would be vertically shifted by a different constant, resulting in branches of the received signal that overlap. At the overlap, one cannot discern which symbol was transmitted.}  


\section{Conclusions}

\added[id=msb]{In this work, I show with mathematical rigour that a linear superposition of chaotic signals with different natural frequencies fully preserves the spectra of Lyapunov exponents and the information content of the source signals. I also show that if each source signal is tuned with appropriated linear coefficients (or power gains), successful decomposition of the source signals and their information content out of the composed signal is possible.} \added[id=m]{Driven by today's huge demand for data, there is a desire to develop wireless communication systems that can handle several sources, each using different frequencies of the spectrum.
 As an application of this wonderful decomposability property that chaotic signals have, I propose} a multi-user and multi-frequency communication system, Wi-C1, where the encoding phase \added[id=m]{(i.e., the preparation of the signal to be transmitted through a physical media)} is based on the correct choice of the linear coefficients, and the decoding phase \added[id=m]{(i.e., the recovery of the transmitted signals and their information content)} is based on the decomposition of the received composed signal.

\added[id=m]{The information encoding capacity of Wi-C1, or the information rate of a signal that is obtained by a linear composition of  chaotic signals, is demonstrated to be equal to the sum of positive Lyapunov exponents of the source signals of each user. If linear coefficients (power gains) are appropriately chosen, and noise is sufficiently low, then the information encoding capacity of Wi-C1 is equal to the information capacity of Wi-C1, or the total rate of information being received/decoded.}

\added[id=msb]{Further improvement for the rate of information could be achieved by adding more transmitters (or receivers) at the expense of reliability. One could also consider similar ideas as in \cite{rosa1997noise,baptista2002information}, which would involve more post-processing, at the expense of weight. Post-processing would involve the resetting of initial conditions in  Eq. (\ref{inverse-map}) all the time, and then using the inverse dynamics up to some specified number of backward iterations to estimate the past of $u^{(k)}_n$. \added[id=m]{One could even think of constructing stochastic resonance detectors to extract the information of a specific user from the received composed signal \cite{Galdi1998EvaluationOS}}. These proposed analyses for the improvement of performance in speed, weight and reliability of the communication are out of the scope of this work. }

\added[id=m]{I have compared the  performance of Wi-C1 with a non-chaotic communication method that is the strongest candidate for the future 5G networks, the Non-Orthogonal Multiplex Access (NOMA), and have shown that Wi-C1 can communicate at higher bit rates for large noise levels in the channel.   

The last Section of this paper is dedicated to conceptually discuss with some mathematical support how chaotic signals and their information content are transformed by physical channels with other non-ideal properties (dispersion and interference with periodic signals), and how this impacts on the multi-user communication system proposed. 
}



\section{acknowledgments}

The author would like to acknowledge preliminary simulations done by Dr Ezequiel Bianco-Martinez with combined maps trajectories, and an email communication by Dr Pedro Juliano Nardelli and Dr Daniel B. da Costa suggesting ``the investigation of Non-Orthogonal Multiple Access (NOMA) in chaos-based communication systems". This communication has motivated the author to introduce the NOMA framework and compare it with the proposed Wi-C1 in \textbf{Sec. VI of SM}. The analysis of  performance (Fig. \ref{paper-figure1c-e}((C))), and discussions involving similarities (and dissimilarities) between the uplink and the downlink scenarios have also benefited from this communication. The author would finally like to thank discussions about detailed numerical solutions of Eqs. (\ref{rescaled-corron})-(\ref{multipath-returnmap-MSB}) performed by MSc Louka Kovatsevits, for arbitrary values of $\tau_l$ and $\beta({\gamma})$. A follow-up paper reporting this analysis will be submitted elsewhere. Points in Fig. 3 of Ref. \cite{YangIEEE2016} were extracted using the software Plot Digitizer. The following references \cite{corron2006chaos,liu2019noise,kizilirmak2016non,benjebbour2017overview,dai2018survey,hampton2013introduction} have contributed to the material 
presented in \textbf{SM}. 


\bibliographystyle{apsrev4-2} 
\bibliography{apssamp}

\section*{Appendix A: The continuous hybrid-dynamics chaotic wavesignal generator}

A wonder of chaotic oscillations for communication is the system proposed in Ref. \cite{corron2010matched}. As  originally proposed, the system is operating in a time frame whose infinitesimal is denoted by $dt$, has a natural frequency $f_0=1$, a natural period $T_0=1$, and an angular frequency $\omega_0=2\pi$. With an appropriate rescaling of time to a new time-frame $dt{^{\prime}}=\gamma dt$, it can be rewritten as in Eq. (\ref{rescaled-corron}), which I reproduce here to simplify understanding of this and following Sections of this Appendix. 
\begin{equation}
    \ddot{x}-2\beta\dot{x}+(\omega^2+\beta^2)(x-s(t))=0, 
\label{rescaled-corron-app}
\end{equation}
\noindent
where $s(t) \in (-1,1)$ is 2-symbols alphabet discrete state that switches value by the signum function $s(t)={x(t)/|x(t)|}$, whenever $|x(t)|<1$ and $\dot{x}=0$. In this new time-frame, the natural frequency is $f(\gamma)=(1/\gamma)$ (angular frequency equals $2\pi f$), the period $T(\gamma)=1/f(\gamma)=\gamma T_0(\gamma=1)$, and $\beta(\gamma) =\beta(\gamma=1) f(\gamma)$, where $0 < \beta(\gamma=1) \leq \ln{(2)}$. $\frac{\beta}{f}$ will be further used to denote $ \frac{\beta(\gamma)}{f(\gamma)} = \beta(\gamma=1)$.

Equation (\ref{rescaled-corron-app}) (and Eq. (\ref{rescaled-corron})) has an analytical solution that links its continuous form to its symbolic encoding, provided by the discrete state $s_n$ obtained by sampling the time 
at $t=n/f$, where $n=\lfloor f t \rfloor$ is the floor function that extracts the integer part of $ft$ \cite{patentUK,patentCHINA,ren2016experimental,yao2017chaos}:
\begin{eqnarray}
x(t)=s_n+\left\{ -s_n  + (1-e^{-\beta/f}) \sum_{i=0}^{\infty} s_{i+n} e^{-i \beta/f} \right\} \nonumber \\
\times e^{\beta (t-nT)}\left( \cos \omega t -\frac{\beta}{\omega} \sin \omega t \right).
\label{analy-sol-corron}
\end{eqnarray}
\noindent
In this equation, $s_n$ represents the binary symbol associated to the time interval $nT \leq t <(n+1)T$, where $s_n=s(t=nT)$. Sampling the time at this same rate a discrete mapping of $x(t)$ can be constructed 
\begin{equation}
x_n=e^{n \beta/f}\left\{x_0 - (1-e^{-\beta / f}) \sum_{i=0}^{n-1} s_i e^{-i \beta /f } \right\}.    
\label{map-sol-corron}
\end{equation}

Moreover, this solution can be written in terms of an infinite sum of basis function whose coefficients are the symbolic encoding of the analogical trajectory ($s_n$). This representation allows for the creation of a matched filter which receives as the input the signal $x(t)$ corrupted by white Gaussian noise (AWGN) and produces as the output an estimation of $x(t)$.   

It offers in a single system all the benefits of both the analogical and digital approaches to communicate. The continuous signal copes with the physical medium, and the digital 
representation provides a translation of the chaotic signal to the digital language that we and machines understand.  Supposing the information to be communicated is a binary stream $\textbf{b}=\{b_0,b_1,b_2,\ldots\}$ a signal can be created (the source encoding phase) such that $s(t) = (2b_n-1)$, for $nT \leq t < (n+1)T$ \cite{patentUK}. The so called source encoding phase is thus based on a digital encoding. Moreover, the discrete variable $s_n$ is the symbolic encoding of the chaotic trajectory in the space $x,\dot{x}$. 

This kind of hybrid chaotic system to communicate is not unique. Corron and Blakely \cite{corron2015chaos}  and Corron, Cooper and Blakely 
\cite{corron2016analytically} have recently proposed other similar chaotic systems to that of Eq. (\ref{rescaled-corron-app}) (and Eq. (\ref{rescaled-corron-app})).    
It was hypothesized in Ref. \cite{corron2015chaos} that the optimal waveform that allows for a stable matched filter is a chaotic waveform. In this work, we provide support for this conjecture, but by showing that stability for the recovery of the information can be cask in terms of the non-existence of negative  Lyapunov exponents of Eq. (\ref{rescaled-corron-app}) (and Eq. (\ref{rescaled-corron-app})). This is shown in section \ref{LE-continous} of this Supplementary Material. Had this system negative LEs, its inverse dynamics - the matched filter - responsible to filter out noise of the received signal would possess a positive LE making it to become unstable to small perturbations in the input of the matched filter (the received signal). 

\section*{Appendix B: The return mapping of the received signal for a single user and an arbitrary number of propagation paths, in the noiseless channel} 

In Eq. (\ref{analy-sol-corron}), $s_n$ represents the binary symbol associated to the time interval $nT \leq t <(n+1)T$, where $s_n=s(t=nT)$.

The received signal in the noiseless wireless channel with a single user can be modelled by  
\begin{equation}
r(t)    = \sum_{l=0}^{L-1} \alpha_l x(t-\tau_l), 
\label{wireless_channel-one-user}
\end{equation}
\noindent
where there are $L$ propagation paths, each with an attenuation factor of $\alpha_l$ and a time-delay $\tau_l$ for the signal to arrive to the receiver along the path $l$ (with $0=\tau_0 < \tau_2 < \ldots < \tau_{L-1}$).

To obtain a map solution for Eq. (\ref{wireless_channel-one-user}), we need to understand which symbol $s_{n^{\prime}}$ is associated to the time.   
$t-\tau_l$. Let us define the time-translated variable 
\begin{equation}
t^{\prime} = t-\tau_l, 
\label{t-prime-with-t}
\end{equation}
\noindent
where $t$ represents the global time for all elements involved in the communication, 
the time that a certain signal $x(t)$ was generated by a user from the chaotic system 
in Eq. (\ref{rescaled-corron-app}), the ``transmitter". $t^{\prime}$ represents the delayed-time. The clock of the user, the ``receiver", is at time $t$ but it receives the signal $r(t^{\prime})$. The receiver decodes for the symbol $s_{n^{\prime}}$, where 
\begin{equation}
    n^{\prime} = \lfloor f t^{\prime} \rfloor. 
\label{n-prime-tau_l}
\end{equation}
The operator $\lfloor \cdot \rfloor$ represents the floor function. 
The transmitter constructs a map at times $t=n/f$, so Eqs. (\ref{t-prime-with-t}) ans (\ref{n-prime-tau_l}) can be written as 
\begin{eqnarray}
    n^{\prime} = \lfloor n - f \tau_l \rfloor = n - \lceil f \tau_l \rceil 
\label{n-prime-tau_l2}, \\
t^{\prime} = \frac{n}{f} - \tau_l, 
\label{t-prime-with-t2}
\end{eqnarray}
where the operator $\lceil \cdot \rceil$ represents the ceiling function.

In the time-frame of $t^{\prime}$, the solution in (\ref{analy-sol-corron}) multiplied by an arbitrary attenuation factor can be written as 
\begin{eqnarray}
\alpha_l x(t^{\prime})&=&\alpha_l s_{n^{\prime}}+ \alpha_l \left\{ -s_{n^{\prime}} + (1-e^{-\beta/f}) \sum_{i=0}^{\infty} s_{i+n^{\prime}} e^{-i \beta/f} \right\} \nonumber \\
&\times& e^{\beta(t^{\prime}-n^{\prime}T)}\left( \cos \omega t^{\prime} -\frac{\beta}{\omega} \sin \omega t^{\prime} \right).
\label{analy-sol-corron-tprime}
\end{eqnarray}

Let us understand what happens to the oscillatory term $\left( \cos \omega t^{\prime} -\frac{\beta}{\omega} \sin \omega t^{\prime} \right)$ in Eq. (\ref{analy-sol-corron-tprime}). Using Eq. (\ref{t-prime-with-t2}), we obtain that 
\begin{equation}
  \cos{(\omega t^{\prime})} -\frac{\beta}{\omega} \sin{(\omega t^{\prime})} = \cos{\left(2\pi \frac{\tau_l}{T}\right)} + \frac{\beta}{\omega} \sin{\left(2\pi \frac{\tau_l}{T}\right)}. 
  \end{equation}

So, 
\begin{equation}
\alpha_l x(t^{\prime})=\alpha_l s_{n^{\prime}}+ \alpha_l \kappa_l \left\{ -s_{n^{\prime}} + (1-e^{-\beta/f}) \sum_{i=0}^{\infty} s_{i+n^{\prime}} e^{-i \beta/f} \right\},  \label{analy-sol-corron-tprime2}
\end{equation}
where 
\begin{equation}
  \kappa_l  = e^{\beta (t^{\prime}-n^{\prime}T)} \left\{ \cos{\left(2\pi \frac{\tau_l}{T}\right)} + \frac{\beta}{\omega} \sin{\left(2\pi \frac{\tau_l}{T}\right)} \right\}.
  \label{kappa}
  \end{equation}

Let us calculate the previously shown quantities for a delayed-time $t^{\prime\prime}$ one period ahead in time of  $t^{\prime}$:

\begin{eqnarray}
t^{\prime\prime} = t^{\prime} + T = t^{\prime} + 1/f \label{tprimeprime}, \\
n^{\prime\prime} = \lfloor f t^{\prime\prime} \rfloor =  1 + n^{\prime}.  \label{nprimeprime}
\end{eqnarray}

It can be also be written that 
\begin{equation}
t^{\prime\prime} - n^{\prime\prime}T =     t^{\prime} - n^{\prime} T.
\label{primeprime-to-prime}
\end{equation}

This equation can be derived by doing $
t^{\prime\prime} - n^{\prime\prime}T = t^{\prime} + \frac{1}{f} -n^{\prime}T -T
$.

Using Eqs. (\ref{t-prime-with-t}), (\ref{tprimeprime}), and (\ref{primeprime-to-prime}), it is possible to obtain   
\begin{eqnarray}
\cos{(\omega t^{\prime\prime})} -\frac{\beta}{\omega} \sin{(\omega t^{\prime\prime})} &=& \cos{(\omega t^{\prime})} -\frac{\beta}{\omega} \sin{(\omega t^{\prime})} \label{sincos-primeprime-to-prime}, \\
&=& \cos{\left(2\pi \frac{\tau_l}{T}\right)} + \frac{\beta}{\omega} \sin{\left(2\pi \frac{\tau_l}{T}\right)},  \nonumber \\
e^{\beta(t^{\prime\prime}-n^{\prime\prime}T)} &=& e^{\beta(t^{\prime}-n^{\prime}T)}. \label{exp-term-primeprime-to-prime}
\end{eqnarray}

So, the attenuated signal at time $t^{\prime\prime}$ is given by 
\begin{eqnarray}
\alpha_l x(t^{\prime\prime}) &=& \alpha_l s_{n^{\prime}+1} - 
\label{analy-sol-corron-tprimeprime}   \\
& & \alpha_l \kappa_l  s_{n^{\prime}+1} + \alpha_l \kappa_l \left\{(1-e^{-\beta/f}) \sum_{i=0}^{\infty} s_{i+n^{\prime}+1} e^{-i \beta/f} \right\}. \nonumber
\end{eqnarray}
Returning to Eq. (\ref{analy-sol-corron-tprime2}), notice that by a manipulation of the terms inside the summation, it can be written as 
\begin{widetext}
\begin{eqnarray}
\alpha_l x(t^{\prime})&=&\alpha_l s_{n^{\prime}} +  
\alpha_l \kappa_l \left\{ -s_{n^{\prime}} + (1-e^{-\beta/f})s_{n^{\prime}} +
(1-e^{-\beta/f}) \sum_{i=0}^{\infty} s_{i+n^{\prime}+1} e^{-(i+1) \beta/f} \right\} \nonumber \\
&=&\alpha_l s_{n^{\prime}} +  
\alpha_l \kappa_l \left\{ -s_{n^{\prime}} e^{-\beta/f}  +
(1-e^{-\beta/f}) \sum_{i=0}^{\infty} s_{i+n^{\prime}+1} e^{-(i+1) \beta/f} \right\}.
\label{analy-sol-corron-tprime3} 
\end{eqnarray}
Multiplying this equation by $e^{\beta/f}$ results in 
\begin{equation}
e^{\beta/f} \alpha_l  x(t^{\prime})= e^{\beta/f} \alpha_l   s_{n^{\prime}} 
- \alpha_l \kappa_l s_{n^{\prime}}   + \alpha_l \kappa_l  \left\{
(1-e^{-\beta/f}) \sum_{i=0}^{\infty} s_{i+n^{\prime}+1} e^{-i\beta/f} \right\}. 
 \label{analy-sol-corron-tprime4}
\end{equation}

\end{widetext}

The received signal at time $t$ and $t+T$ are then given by 
\begin{eqnarray}
r(t)    &=& \sum_{l=0}^{L-1} \alpha_l x(t^{\prime}), \\
r(t+T)    &=& \sum_{l=0}^{L-1} \alpha_l x(t^{\prime\prime}). 
\end{eqnarray}
If we observe the received signal only at discrete times $t=nT$, and defining the discrete variable 
$r(nT) \equiv r_n$, we obtain that

\begin{eqnarray}
e^{\beta/f} r_n &=& \sum_{l=0}^{L-1}  \alpha_l \left\{ e^{\beta/f} s_{n^{\prime}} -\mathcal{K}_l  s_{n^{\prime}} 
+ \mathcal{K}_l \mathcal{A} \right\} \label{received-prime},\\
r_{n+1} &=& \sum_{l=0}^{L-1}  \alpha_l \left\{s_{n^{\prime}+1} -\mathcal{K}_l  s_{n^{\prime}+1} 
+ \mathcal{K}_l \mathcal{A} \right\},
\label{received-primeprime}
\end{eqnarray}
\noindent
where 
\begin{equation}
    \mathcal{K}_l = e^{-\beta (\tau_l - \lceil \tau_l/T \rceil T)}\left\{ \cos{\left(2\pi\frac{\tau_l}{T}\right)} + \frac{\beta}{\omega}\sin{\left(2\pi \frac{\tau_l}{T}\right)}\right\}. 
\end{equation}

The variable $\mathcal{K}_l$ is derived by noticing that 
\begin{equation}
    e^{-\beta(\tau_l - \lceil \tau_l/T \rceil T)} = e^{\beta(t^{\prime} - n^{\prime}T)}.
\end{equation}

Comparing Eqs. (\ref{received-prime}) and (\ref{received-primeprime}), we finally arrive at a return map for the received signal with multipath propagation 
\begin{equation}
    r_{n+1} = e^{\beta/f} r_n - \sum_{l=0}^{L-1}  \alpha_l \left( 
    e^{\beta/f} s_{n^{\prime}} - \mathcal{K}_l s_{n^{\prime}} - 
    s_{n^{\prime}+1} + s_{n^{\prime}+1} \mathcal{K}_l \right).
\end{equation}

\section*{Appendix C: The equivalence principle for flows, and the preservation of the Lyapunov Exponents for linearly composed chaotic signals}

Let us assume that information is being encoded by using the R\"ossler attractor. 
User 1 encodes its information in the variable $x_1(t)$ and user 2 in the variable $x_2(t)$. User 2 has a base frequency $Q$ times the one from user 1. Notice that $Q=1/\gamma$, where $\gamma$ is the time-rescaling factor. 
User 1 chaotic signal $x_1$ is generated by
\begin{eqnarray}
\dot{x}_1=-y_1-z_1(t), \nonumber \\
\dot{y}_1=x_1+ay_1, \label{user1-rossler} \\
\dot{z}_1 = b + z_1(x_1-c) \nonumber. 
\end{eqnarray}
User 2 uses the signal $x_2$ generated by : 
\begin{eqnarray}
\dot{x}_2=Q[-y_2-z_2(t)], \nonumber \\
\dot{y}_2=Q[x_2+ay_2], \label{user2-rossler} \\
\dot{z}_2 = Q[b + z_2(x_2-c)] \nonumber,  
\end{eqnarray}
\noindent
where $a$, $b$ and $c$ represent the usual parameters of the R\"ossler attractors. Notice however that this demonstration would be valid to any nonlinear system, the R\"ossler was chosen simply to make the following calculation straightforward to follow. The system of equations in (\ref{user2-rossler}) are already in the transformed time-frame, so that user 2 has a basis frequency $Q$ times larger than user 1. 

The transmitted composed signal can be represented by 
\begin{equation}
O(t) = \alpha_1 x_1(t) + \alpha_2 x_2(t), \label{transmitted-rossler}   
\end{equation}
\noindent
where $\alpha_1$ and $\alpha_2$ are attenuation, or power gain factors. 


The interest is to derive a systems of ODEs that describes all variables involved in ODE system describing the received signal  
\begin{equation}
\dot{O}(t) = \alpha_1 \dot{x}_1(t) + \alpha_2 \dot{x}_2(t). \label{transmitted-rossler-dot}   
\end{equation}
\noindent

Determinism in chaos allows us to write that at a time $t$ there exists a $\tau$ such that $
x_2(t) = x_1(t-\tau)$. More generally, if user 2 has a basic frequency $Q$ times that of user 1, its trajectory at time $t+n\delta t$ ($n \in \mathbb{N}$) can be written in terms of the user 1's trajectory by 
\begin{equation}
x_2(t + n \delta t) = x_1(t + n Q \delta t - \tau).    
\end{equation}

I now define a new set of transformation 
variables given by 
\begin{eqnarray}
w_n^{1x}(t) & = & x_1(t-Q\ n \ \delta t), \label{trans1} \\
w_n^{1y}(t) & = & y_1(t-Q\ n \ \delta t), \label{trans2} \\
w_n^{1z}(t) & = & z_1(t-Q\ n \ \delta t), \label{trans3}
\end{eqnarray}
\noindent
where 
\begin{eqnarray}
    n=\{0,\ldots, N \}, \\
    N \delta t = \tau. 
\end{eqnarray}
Defining the vectors $\mathbf{X}_1(t)=\{x_1(t),y_1(t),z_1(t)\}$ and $\mathbf{W}^1_n(t)=\{w_n^{1x}(t),w_n^{1y}(t),w_n^{1z}(t)\}$, I can write Eqs. (\ref{trans1})-(\ref{trans3}) in a compact form
\begin{equation}
\mathbf{W}^1_n(t) = \mathbf{X}_1(t - Q\ n \ \delta t).    
\end{equation}
Notice also that by defining the vector $\mathbf{X}_2(t)=\{x_2(t),y_2(t),z_2(t)\}$, I can write that at time t
\begin{eqnarray}
\mathbf{W}^1_n(t) &=& 
\mathbf{X}_1(t -\tau + (N - Q\ n) \delta t), 
\label{trans-vector-x1} \\
&=& 
 \mathbf{X}_2\left(t + \frac{(N-Qn)}{Q}\delta t\right). 
\label{trans-vector-x2} \\
\end{eqnarray}
To facilitate the following calculations, I express some terms of Eqs. (\ref{trans-vector-x1}) and 
(\ref{trans-vector-x2}) along the variables $x_1(t)$ and $x_2(t)$ for $n=\{0,1,2,\ldots,N-1,N\}$:
\begin{eqnarray}
w_N^{1x}(t)&=&x_1(t-\tau)=x_2(t), \\
w_{N-1}^{1x}(t)&=&x_1(t-\tau+Q\delta t) = x_2(t+\delta t), \\
\vdots \nonumber  \\
w_2^{1x}(t)&=&x_1(t-2Q \delta t), \\
&=& x_2(t+  \tau/Q - 2\delta t), \nonumber  \\
w_1^{1x}(t)&=&x_1(t-2\delta t), \\
&=& x_2(t+  \tau/Q - \delta t), \nonumber  \\
w_0^{1x}(t)&=&x_1(t) = x_2(t+  \tau/Q). 
\end{eqnarray}
I express the transformation 
variables considering a small displacement  
$Q \delta t$ in time from the time $t$: 
\begin{eqnarray}
\mathbf{W}^1_n(t+ Q \delta t) &=& 
\mathbf{X}_1(t -\tau + (N - Q\ n) \delta t + Q \delta t) 
\label{trans-vector-x1-Deltat} \\
&=& 
 \mathbf{X}_2\left(t + \frac{(N-Qn)}{Q}\delta t + Q \delta t \right).
\label{trans-vector-x2-Deltat} 
\end{eqnarray}
Then, time derivatives can now be defined by  
\begin{eqnarray}
\dot{\mathbf{W}}^1_{n-1}(t) &=& \frac{(\mathbf{W}^1_{n-1} - \mathbf{W}^1_n(t))}{Q\delta t} 
\label{derivadas1}, \\
&& n \in [2,\ldots,N]. \nonumber
\end{eqnarray}

For the variables of the user 2, we have that 
\begin{eqnarray}
\dot{x}_2(t) &=& \frac{x_2(t+\delta t) - x_2(t)}{\delta t}=    \frac{w^2_{N-1}(t) - w^2_{N-1}(t)}{\delta t},\\ 
& = & \dot{w}_N^2(t) \nonumber, 
\end{eqnarray}
\noindent
which takes us to 
\begin{eqnarray}
\dot{\mathbf{W}}^2_n(t) &=& \frac{(\mathbf{W}^2_{n-1} - \mathbf{W}^2_n(t))}{\delta t} 
\label{derivadas2}, \\
&& n \in [1,\ldots,N-1]. \nonumber
\end{eqnarray}

The original variables of the R\"ossler system for user 1 can be written in terms of the new transformed variables by 
\begin{eqnarray}
\mathbf{W}^1_0 & = & \mathbf{X}_1(t), \\
\mathbf{\dot{W}}^1_0 & = & \mathbf{\dot{X}}_1(t), \nonumber
\end{eqnarray}
and for the user 2
\begin{eqnarray}
\mathbf{W}^2_N & = & \mathbf{X}_2(t), \\
\mathbf{\dot{W}}^2_N & = & \mathbf{\dot{X}}_2(t). \nonumber
\end{eqnarray}

Before I proceed, it is helpful to do some considerations, regarding this transformation of variables. Notice that $\dot{x}_2=\frac{w_{N-1}(t) - w_N(t)}{\delta t}$ and $\dot{x}_1=\frac{w_{0}(t) - w_1(t)}{Q \delta t}$. So, $\dot{x}_2 = Q\dot{x}_1$. Moreover, $\mathbf{W}^1_n(t=\mathbf{X}_1(t-\tau+(N-Qn)\delta t))$ and $\mathbf{W}^2_n(t=\mathbf{X}_2(t+(N-Qn)/Q \delta t))$. So, $\mathbf{W}_n^1(t) = \mathbf{W}_n^2(t)$, but their derivatives are not equal. 

In the new variables, the ODE system describing the received signal from user 1 is described by  
\begin{eqnarray}
\dot{w}^{1x}_0=-w^{1y}_0 - w^{1z}_0, \nonumber \\
\dot{w}^{1y}_0=w^{1x}_0 + aw^{1y}_0, \label{user1-rossler-newvariables} \\
\dot{w}^{1z}_0 = b + w^{1z}_0(w^{1x}_0-c) \nonumber,  
\end{eqnarray}
\noindent
and user 2 is described by: 
\begin{eqnarray}
\dot{w}^{2x}_N=-w^{2y}_N - w^{2z}_N, \nonumber \\
\dot{w}^{2y}_N=w^{2x}_N + aw^{2y}_N, \label{user2-rossler-newvariables} \\
\dot{w}^{2z}_N = b + w^{2z}_N(w^{2x}_N-c). \nonumber  
\end{eqnarray}
\noindent

The received signal is described by 
\begin{equation}
O(t)=\alpha_1 w_0^{1x}+ \alpha_2 w_N^{2x}, 
\end{equation}
and its first time derivative 
\begin{eqnarray}
\dot{O}(t)&=&\alpha_1 \dot{w}_0^{1x}+ \alpha_2 \dot{w}_N^{2x}, \nonumber \\
&=& \alpha_1 (-w^{1y}_0 - w^{1z}_0) + \alpha_2 (-w^{2y}_N - w^{2z}_N) \label{received-derivative-new}. 
\end{eqnarray}

A final equation is needed to describe the first time derivative of $\dot{w}_0^{2x}(t)$ in terms of the previously defined new variables. We have that 
\begin{equation}
\dot{w}_0^{2x} = \frac{1}{\delta t}(w_0^{2x}(t+\delta t) - w_0^{2x}(t)).  
\label{derivaw_02x}
\end{equation}
\noindent 
Moreover, 
\begin{eqnarray}
w_0^{2x}(t+\delta t) &=& w_0^{1x}(t+ Q \delta t), \label{derivaw_02x1} \\
w_0^{1x}(t+ Q \delta t) & =& w_0^{1x}(t) + \dot{w}_0^{1x}Q \delta t. \label{derivaw_02x2}
\end{eqnarray}
Placing Eqs. (\ref{derivaw_02x1}) and (\ref{derivaw_02x2}) in Eq. (\ref{derivaw_02x}), takes us to 
\begin{equation}
\dot{w}_0^{2x} = Q \dot{w}_0^{1x} = \frac{1}{\delta t} (w_0^{1x} - w_1^{1x}).     
\label{derivaw_02x-final}
\end{equation}

The variational equations of the systems formed by Eqs. (\ref{derivadas1}), (\ref{derivadas2}),  (\ref{user1-rossler-newvariables}), (\ref{user2-rossler-newvariables}), (\ref{received-derivative-new}), and (\ref{derivaw_02x-final}) can be constructed by defining the perturbed variables $\tilde{w}^{jk}_i=w^{jk}_i + \delta w^{jk}_i$, with the index representing $j \in \{1,2,3\}$, $k \in \{ x,y,z \}$, $i \in \{0,\ldots,N\}$, whose first 
derivative is $\dot{\tilde{w}}^{jk}_i=\dot{w}^{jk}_i + \delta \dot{w}^{jk}_i$. Also, $\tilde{O}=O+\delta O$, for the received signal. The Jacobian matrix of the variational equations is thus given by 

\begin{widetext}
$
\begin{array}{ccccccccccccccc}
 & \delta w_0^{1x} & \delta w_0^{1y} & \delta w_0^{1z} & \delta w_1^{1x} & \delta w_2^{1x}  & \hdots &  \delta w_0^{2x} & \delta w_1^{2x} & \delta w_2^{2x} & \hdots & \delta w_N^{2x} & \delta w_N^{2y} & \delta w_N^{2z} & \delta O \\
  \delta {\dot{w}^{1x}_0} &. & -1 & -1 &. &. &. &.  &. &. &. &. &. &.&. \\
     \delta {\dot{w}^{1y}_0} & 1 & a &. &. &. &. &. &.  &. &. &. &. &. &. \\
       \delta {\dot{w}^{1z}_0} & w_0^{1z} & . & w_0^{1x} - c&. &. &. &. &. &. &. &. &. &. &. \\
         \delta {\dot{w}^{1x}_1} &. &. &. & (Q\delta t)^{-1} & -(Q\delta t)^{-1}& .&. &. &. &. &. &. &. &. \\
           \delta {\dot{w}^{1x}_2} &. &. &. &. & (Q\delta t)^{-1}  &. &. &. &. &. &. &. &. &. \\
             \vdots &. &. &. &. &. &. & \ddots & . &. &. &.  &. &. & \vdots \\
              \delta {\dot{w}^{2x}_0} & \delta t^{-1} &. &. &  - \delta t^{-1} &. &. &. &. &. &. &. &. &. &. \\
               \delta {\dot{w}^{2x}_1} &. &. &. &. &. &. & \delta t^{-1} & -\delta t^{-1} &. &. &. &. &. &. \\
                 \delta {\dot{w}^{2x}_2} &. &. &. &. &. &. &. & \delta t^{-1} & -\delta t^{-1} &. &. &. &. &. \\
            \vdots &. &. &. &. &. &. &. &. &. & \ddots &. &. &. & \vdots \\
             \delta {\dot{w}^{2x}_N} &. &. &. &. &. &. &. &. &. &. &. & -Q & -Q &. \\
              \delta {\dot{w}^{2y}_N} &. &. &. &. &. &. &. &. &. &. & Q & Qa &. &. \\
               \delta {\dot{w}^{2z}_N} &. &. &. &. &. &. &. &. &. &. &Qw_N^{2z} &. & Q(w_N^{2x} - c) &. \\
                 \delta \dot{o} &. & -\alpha_1 & \alpha_1 &. &. &. &. &. &. &. &. & -Q \alpha_2 & -Q \alpha_2 &. \\
        
\end{array}
$
\end{widetext}

The upper-left diagonal block 
\\

$\begin{pmatrix}
.& -1 & -1 \\
1 & a &. \\
w_0^{1z} & . & w_0^{1x} - c
\end{pmatrix}$, 

\noindent
is responsible to produce the same 3 Lyapunov exponents $\chi_1,\chi_2=0,\chi_3$ ($\chi_1>0,\chi_3<0 $) of the R\"ossler attractor, for the user 1. 

The bottom-right diagonal block 

$\begin{pmatrix}
.& -Q & -Q \\
Q & Qa &. \\
Qw_0^{1z} & . & Q(w_0^{1x} - c) 
\end{pmatrix}$,

\noindent
is responsible to produce the 3 Lyapunov exponents $Q\chi_1,\chi_2=0,Q\chi_3$ (so Q times the Lyapunov exponents of the R\"ossler attractor), for the user 2. The diagonal elements of the Jacobian will produce 
$N-1$ Lyapunov exponents equal to $(Q \delta t)^{-1}$ and $N-1$ exponents $-\delta t^{-1}$. The signs of 
the "infinities" Lyapunov exponents are a consequence of the way the derivatives were defined, and they could have been made to have the same signs. These exponents represent a higher-dimensional dynamics that effectively does not participate in the low-dimensional ordinary dynamics of the measured received signal. They are a consequence of the transformation of a time-delayed system of differential equations into an ODE, without explicitly time dependence.       

Concluding, the spectra of Lyapunov exponents of the dynamics generating the signals for user 1 and 2 are preserved in the received signal, and are not affected by the combination of the signals.

\end{document}



\title{Supplementary Material of paper ``Wonders of chaos for communication"}

\author{Murilo S. Baptista}
 \affiliation{Institute for Complex Systems and Mathematical Biology, University of Aberdeen, AB24 3UX, Aberdeen, UK.}
\maketitle

\section{Calculation of the Lyapunov exponents of the continuous hybrid system}\label{LE-continous}

I now proceed to estimate the Lyapunov exponents (LEs) of the continuous hybrid chaotic system: 
\begin{equation}
    \ddot{x}-2\beta\dot{x}+(\omega^2+\beta^2)(x-s(t))=0, 
\label{rescaled-corron-lyap}
\end{equation}
\noindent
where $s(t) \in (-1,1)$ is 2-symbols alphabet discrete state that switches value by $s(t)={x(t)/|x(t)|}$, whenever $|x(t)|<1$ and $\dot{x}=0$. In this new time-frame, the natural frequency will be $f=(1/\gamma)f_0$ (angular frequency equals $2\pi f$), the period $T=\gamma T_0=1/f$, and $\beta \leq f \theta $ and $0 < \theta \leq \ln{(2)}$.  

Firstly, I transform the hybrid system so that it is fully described by a set of first order Ordinary Differential Equations (ODEs). In order to do so, I first define the variable $y(t)=\dot{x(t)}$. Then, notice that the discrete variable assumes either -1 or 1 values, and the switch to either one of the values happen when $y=0$. So, $\dot{s}=0$, except when $\dot{x}=0$. Thus, $\dot{s}$ is a function of the 2 variables $(x,y)$. If we were able to find a continuous description of 
$s$, its first-time derivative would be described by 
$\dot{s}=\frac{\partial s}{\partial x} \frac{dx}{dt} + \frac{\partial s}{\partial y} \frac{dy}{dt}$. Making the approximation that $s$ only depends on $x$, such that it is described only by the signum function $s(x)=\frac{x}{|x|}$, we would have that $\dot{s} = 2 y \delta(x)$,
if $y=0$ or $\dot{s} = 0$, otherwise, $y \not= 0$,  
where $\delta(x)$ is the delta's Dirac function.  Notice also that 
 $\frac{\partial s}{\partial x}$ is unlikely to assume a value different than zero, since when $s$ switches values from -1 to 1 (or from 1 to -1) $x>0$ ($x<0$).   

After these considerations, it is possible to write an effective system of ODEs for the Hybrid system as
\begin{eqnarray}
\dot{x} &=& y, \nonumber \\
\dot{y} &=& 2 \beta y - (\omega^2+\beta^2)(x-s), \label{first-order-equivalent} \\
\dot{s} &=& 0. \nonumber 
\end{eqnarray}

When calculating the Lyapunov exponents of a system of ODEs, the interest lies with the variational equations of the system that describe how perturbations propagate along the trajectory. Defining the vector $\bm{Y}(t)=(x(t),y(t),s(t))$, a perturbed trajectory $\bm{X}(t)$ can be described by $\bm{X}(t) = \bm{Y}(t) + \bm{\delta(t)}$. Then, the ODE system describing how perturbations propagate is then given by the variational equation 
\begin{equation}
\bm{\dot{\delta}(t)} = \bm{J}  \bm{\delta(t)},
\label{perturbation-propation}
\end{equation}
\noindent
where $\bm{J}$ represents the Jacobi matrix of partial derivatives of the vector flow in Eq. (\ref{first-order-equivalent}), and it is given by  
\begin{eqnarray}
\bm{J} &=& 
\begin{pmatrix}
0  & 1 & 0 \\
-(\omega^2 + \beta^2) & 2\beta & (\omega^2+\beta^2) \\
0 & 0 & 0
\end{pmatrix}.
\end{eqnarray}
This is consistent with the idea that there is no propagation of perturbations along the direction of the variable $s$, once the system is a forced one, and defined as to have $s$ as either -1 or 1. Moreover, 
any continuous system must possess a null LE.

Notice also that whereas $\dot{y}$ is affected by, the propagation of perturbations does not depend on whether $s=1$ or $s=-1$. The interest thus is to calculate how perturbations propagate along the plan $(x,y)$, which would mean that $\bm{\delta} \in \Re^2$ and  
\begin{eqnarray}
\bm{J} &=& 
\begin{pmatrix}
0  & 1  \\
-(\omega^2 + \beta^2) & 2\beta 
\end{pmatrix}.
\end{eqnarray}
 I then search for a general solution for $\bm{\delta}(t)$ as 
\begin{equation}
 \bm{\delta}(t) = \bm{\Psi}(t) \bm{v},  
\end{equation}
\noindent
where $\bm{v}$ is a constant initial perturbation vector, and $\bm{\Psi}(t)$ is the fundamental solution matrix of the system in Eq. (\ref{perturbation-propation}) whose columns are independent solutions 
of this system of equations. 

The 1-dimensional Lyapunov exponent of the system (\ref{first-order-equivalent}) in the direction of 
$\bm{v}$ along a trajectory with initial condition $\bm{Y}_0$ is given by 
\begin{equation}
\chi(\bm{Y}_0, \bm{v}) = \lim_{t\rightarrow \infty} \frac{1}{t} \ln{||\Psi(t,\bm{Y}_0) \cdot \bm{v}||}.     
\label{1d-le}
\end{equation}

The spectra of LEs can be obtained by setting $\bm{v}$ to be equal to an eigenvalue $\bm{w}_i$ of the matrix $\Psi(t,\bm{Y}_0)^T \Psi(t,\bm{Y}_0)$. More specifically \cite{maria2019}, Eq. (\ref{1d-le}) is algebraically equal to
\begin{equation}
\chi(\bm{Y}_0, \bm{v}) = \lim_{t\rightarrow \infty} \frac{1}{2t} \ln{[\bm{v}^T \cdot \Psi^T \cdot (t,\bm{Y}_0) \cdot \Psi(t,\bm{Y}_0) \cdot \bm{v}]}.     
\label{1d-le1}
\end{equation}
And, the LE in the direction of $\bm{w}_i$ ($\bm{v}=\bm{w}_i$) can then be calculated by 
\begin{eqnarray}
\chi(\bm{Y}_0, \bm{w}_i) &=& \lim_{t\rightarrow \infty} \frac{1}{2t} \ln{[\bm{w}_i^T \cdot \Psi^T \cdot (t,\bm{Y}_0) \cdot \Psi(t,\bm{Y}_0) \cdot \bm{w}_i]}, \nonumber \\
& = & \lim_{t\rightarrow \infty} \frac{1}{2t} \ln{[\bm{w}_i^T \cdot \Lambda_i \cdot \bm{w}_i]}, \nonumber \\
 & = & \lim_{t\rightarrow \infty} \frac{1}{2t} \ln{(\Lambda_i)} \label{LE-continuous}, 
\end{eqnarray}
\noindent
since $\frac{1}{2t}\ln{||\bm{w}_i||^2} \rightarrow 0$, as $t\rightarrow \infty$. Oseledec's multiplicative ergodic theorem \cite{eckmann1985ergodic} guarantees that the limit in Eq. (\ref{LE-continuous}) exists and moreover it does not depend on the initial condition, $\bm{Y}_0$, for typical initial conditions.    
It is now clear to see that all that matters to calculate the LE is the matrix $\Psi(t,\bm{Y}_0)$. 
Typically, all that matters are the eigenvalues $\Lambda_i$ of 
$\Psi(t,\bm{Y}_0)^T \Psi(t,\bm{Y}_0)$, but I will show that for this system, it is only required the knowledge of $\Psi(t,\bm{Y}_0)$.  

The fundamental solution matrix $\Psi(t,\bm{Y}_0)$ is given by 
\begin{equation}
\bm{\Psi} = e^{\beta t} \bm{V}(t,V_1,V_2,k_1,k_2), 
\end{equation}
where $V(t,V_1,V_2)$ is a time oscillatory matrix function ($V(t,V_1,V_2,k_1,k_2) \in \Re^2$) that is bounded, and also a function of the components ($V_1$, $V_2$) of the complex eigenvectors of $\bm{J}$, and $k_1$ and $k_2$ are constants set by the initial value. 

Any unitary initial perturbation vector $\bm{v}$ chosen on the plan $(x,y)$ will by Eq. (\ref{1d-le}) grow its magnitude by $e^{\beta t}$. The matrix $\bm{V}(t,V_1,V_2,k_1,k_2)$ will only be responsible for its rotation. So, Eq. (\ref{1d-le}) produces the same degenerated LE equal to 
\begin{equation}
\chi = \beta.     
\label{1d-le-forward}
\end{equation}

\section{The equivalence principle for maps}

To understand how to use the equivalent principle to derive general returning map expressions to the signal arriving at the BS, I make the assumption that there is one user with a basic frequency $f^{(1)}=f$, and then a second user with a frequency $f^{(2)}=2f$. Using the equivalence principle, the combined signal arriving at the BS can be written as 
\begin{equation}
O_{n}=\tilde{\gamma}^{(1)}u^{(1)}_n +   \tilde{\gamma}^{(2)}u^{(1)}_{n-f^{(2)}\tau}. 
\label{combined_received-general}
\end{equation}
We now transform the delay system into an ordinary systems of equations, by 
defining a new set of $\tau+1$ variables
\begin{equation}
W_n^{(T)} = u^{(1)}_{n-f^{(2)}T}, \mbox{\ \ } T=\{0,1,\ldots,\tau\}. 
\label{new-variables}
\end{equation}
Noticing that $W_n^{(\tau)} = u^{(1)}_{n-f^{(2)}\tau}=u_n^{(2)}$, we can rewrite Eq. (\ref{combined_received-general})
as 
\begin{equation}
O_{n} = \tilde{\gamma}^{(1)}u^{(1)}_n +   \tilde{\gamma}^{(2)} W_n^{(\tau)}, 
\label{combined_received-general1}
\end{equation}
\noindent
and making 1 iteration in Eq. (\ref{combined_received-general1}), using Eqs. (\ref{user1-shift}) and (\ref{user2-shift})
\begin{eqnarray}
u^{(1)}_{n+1}=2u^{(1)}_n-\lfloor 2u^{(1)}_n \rfloor \equiv 2u^{(1)}_n- b^{(1)}_n \label{user1-shift}, \\
u^{(2)}_{n+1}=4u^{(2)}_n-\lfloor 4u^{(2)}_n \rfloor \equiv 4u^{(2)}_n- b^{(2)}_n, \label{user2-shift}
\end{eqnarray}
\noindent
and knowing that $W_{n+1}^{(\tau)}  = 2^{f^{(2)}} W_n^{(\tau)} - \lfloor 2^{f^{(2)}} W_n^{(\tau)}  \rfloor = 2^{f^{(2)}} W_n^{(\tau)} - \lfloor 2^{f^{(2)}} u_n^{(2)}  \rfloor = 2^{f^{(2)}} W_n^{(\tau)} - b_n^{(2)}$ we obtain that 
\begin{eqnarray}
O_{n+1} &=& 2 \tilde{\gamma}^{(1)} u^{(1)}_{n} - \tilde{\gamma}^{(1)} b_n^{(1)} + 
2^{f^{(2)}}\tilde{\gamma}^{(2)} W_n^{(\tau)} \\
&-& \tilde{\gamma}^{(2)} b_n^{(2)} \nonumber
\label{combined_received-general-n1}. 
\end{eqnarray}

Comparing Eq. (\ref{combined_received-general-n1}) with Eq. (\ref{combined_received-general1}), we finally obtain a general expression for the return map  
\begin{eqnarray}
    O_{n+1} &=& 2^{f^{(2)}} O_n - (2^{f^{(2)}} - 2) \tilde{\gamma}^{(1)} u^{(1)}_{n} -   \tilde{\gamma}^{(1)} b_n^{(1)} \\ 
    &-& 
\tilde{\gamma}^{(2)} b_n^{(2)} \nonumber
\label{combined_received-general-return}. 
\end{eqnarray}
If the two users have the same frequency, then $f^{(2)}=f^{(1)}=1$, then $ O_{n+1} = 2 O_n -   \tilde{\gamma}^{(1)} b_n^{(1)} - 
\tilde{\gamma}^{(2)} b_n^{(2)}$.

\section{Power gains in the downlink configuration}

Unlike the uplink, in which each user $k$ decides on the power gain $\gamma^{(k)}$ based on their knowledge of the attenuation factor $\alpha^{(k)}_l$, in the downlink, the operator at the BS must decide on a unique value for power gain $\gamma^{(k)}$. The decision is done in such a way to compensate for the channel with the largest attenuation factor (or minimal value of $\alpha_l^{(k)}$). 
\begin{equation}
\gamma^{(k)} \equiv \gamma^{*} = \frac{1}{\min_{\forall i}{[\alpha_0^{(i)}]}}.     
\label{gamma-downlink}
\end{equation}
Neglecting multipath and choosing for the uplink configuration that $\gamma^{(k)}=[\alpha_0^{(k)}]^{-1}$ and for the downlink configuration as in Eq. (\ref{gamma-downlink}),  we arrive that 

\begin{eqnarray}
O(t)_{u}    &=& \sum_{k=1}^N  \tilde{\gamma}^{(k)} x^{(k)}(t) + w(t) \label{WiChaos_up-no-mp1}\\
O^{(m)}(t)_{d}  &=&   \alpha_0^{(m)} \gamma^{*}  \sum_{k=1}^N \tilde{\gamma}^{(k)} x^{(k)}(t),  
+ w^{(m)}(t) \label{WiChaos_down-no-mp1}
\end{eqnarray}
\noindent
and so if   $\alpha_0^{(m)} = \min_{\forall i}{[\alpha_0^{(i)}}]$, then equation for the received uplink 
signal $O(t)_{u}$ is equal to the received signal in the downlink $O^{(m)}(t)_{d}$, except by the noise 
term. Otherwise, these 2 equations differ additionally by a constant factor ($\alpha_0^{(m)} \gamma^{*}$) which will only actually result in that effectively the signal to power rate will change. 

\section{The matched filter decoding approach}

A more sophisticated approach to decode information is based on a matched filter \cite{corron2010matched}. In here, I show that the system formed by Eq. (\ref{rescaled-corron}) 
\begin{equation}
    \ddot{x}-2\beta\dot{x}+(\omega^2+\beta^2)(x-s(t))=0, 
\label{rescaled-corron}
\end{equation}
\noindent
and its matched filter can be roughly approximately described by the  unfolded Baker's map, a result that allows us to understand that the decoding of a message by a user from the combined signal solely depends on the inverse dynamics of this user. If the dynamical equations generating the chaotic signal to be transmitted (in this case Eq. (\ref{rescaled-corron})) possesses no negative Lyapunov exponents (demonstrated in Sec. VIII of this \textbf{SM}), then attractor estimation of a noisily corrupted signal can be done using its autonomous time-inverse dynamics that is stable and possess no positive LEs (\textbf{shown in Sec. IX of SM}. The  evolution to the 
future of the time-inverse dynamics 
is described by a system of ODE hybrid equations obtained by the time-rescaling $d/dt^{\prime} = -d/dt$ applied to 
Eq. (\ref{rescaled-corron}) resulting in  
\begin{equation}
    \ddot{y}+2\beta\dot{y}+(\omega^2+\beta^2)[y-\eta(t)]=0, 
\label{rescaled-corron-inverse}
\end{equation}
\noindent
\noindent
where the variable $y$ represents the $x$ in time-reverse. 

Let us define the symbol $S(t)$ to be a discrete encoding of 
$y(t)$, by the same rules that $s(t)$ is calculated from $x(t)$. So, $S(t) \in (-1,1)$ is a 2-symbols alphabet discrete state that switches value by the signum function $S(t)={y(t)/|y(t)|}$, whenever $|y(t)|<1$ and $\dot{x}=0$.

Let me do some introductory remarks, regarding  
Eqs. (\ref{rescaled-corron-inverse}) and  
($\ref{rescaled-corron}$). 
There are numerous ways to set the initial conditions of Eq. (\ref{rescaled-corron-inverse}), and depending on how that is done, one will obtain different results. If at time 
$t$ we set as an initial condition in Eq. (\ref{rescaled-corron-inverse}) that $y(t)=x(t+T)$, then at the time $t+T$ it could be true that $y(t+T)=x(t)$, if other initial conditions are appropriately set. 
Notice that if $y(t+T)=x(t)$, then the symbol $S(t+T)$ encoding 
$y(t+T)$ should be a match for the symbol $s(t)$ encoding 
$x(t)$. 


Let us now understand the dynamical mechanism responsible to allow partial reconstruction of the attractor produced by Eq. (\ref{rescaled-corron}) by Eq. (\ref{rescaled-corron-inverse}), which allows for the decoding of the message using the inverse dynamics. Since Eq. (\ref{rescaled-corron-inverse}) has no positive Lyapunov Exponents, if there is ADWN in the variable $x(t)$, i.e., $x(t+T)+\xi(t+T)$, and we set in Eq. (\ref{rescaled-corron-inverse}) that
$y(t)=x(t+T)+\xi(t+T)$, then at the time $t+T$ we could have that 
$y(t+T) = x(t) + \epsilon$, more specifically that $(y(t+T) - x(t)) e^{\chi T}  \propto  \xi(t+T)$.  

Any corrupted variable or set of variables that is transmitted can be plugged into the inverse dynamics in Eq. (\ref{rescaled-corron-inverse}) in order to estimate a delayed version of the transmitted signal. In practice, we have not too many options. The transmitted variable is $x(t)$. This variable alone, however, might not contain sufficient information for the inverse dynamics to estimate the past of all the variables in Eq. (\ref{rescaled-corron}). This is because the attractor of Eq. (\ref{rescaled-corron}) lives effectively in two subspaces $s(t)=1$ and $s(t)=-1$. To be able to determine what will be the past of the variable $x(t)$, one might need to additionally have a good estimation of either the values of $\dot{x}$, or the 
values of the discrete variable, basically the message to be transmitted. First-time derivative variables $\dot{x}$ corrupted by noise are typically non-optimal to signal processing. So, we remain with that to use the inverse dynamics in Eq. (\ref{rescaled-corron-inverse}) a good estimation of either $x(t)$ or $s(t)$ (or both) are required.  Then, driving the inverse dynamics with that estimated variable should provide a good estimation for the short-term past of the variable $x(t)$.

System (\ref{rescaled-corron-inverse}) was originally derived in \cite{corron2010matched}, 
but using approaches from signal analysis 
by searching for a matched filter for the signal generated by 
Eq. (\ref{rescaled-corron}) that maximises signal to noise rate.
In that work, a first order filter (also called the auxiliary variable) was defined by the first-time derivative of the variable $\eta(t)$ in the inverse dynamics of 
Eq. (\ref{rescaled-corron-inverse}) by
\begin{equation}
    \dot{\eta}(t) = \tilde{x}(t+T)-\tilde{x}(t), 
\label{first-order-filter}
\end{equation}
\noindent
where $\tilde{x}(t)=x(t)+\xi(t)$. 

The use of this matched filter to communication channels possessing multipath propagation was done in a series of papers \cite{ren2017chaotic, bai2018chaos,yao2017chaos}.
Estimation of the symbolic sequence can also be done directly from $x(t)$ \cite{liu2019noise} 
by noticing that if $s_n=1$ ($s_n=-1$) then
$\int_{t}^{t+T} \tilde{x}(t) dt>0$ ($\int_{t}^{t+T} \tilde{x}(t) dt < 0$). These works have in common that their  
approach to decoding is based on a good estimation of the discrete variable $s(t)$. 

The filter was derived using arguments of signal analysis, but its working dynamical mechanism can be understood if one notices that $\dot{\eta}$ as defined represents a sum of approximate delta functions whose integral are approximately unitary and therefore   
$\eta(t) = s(0) + \int_0^{t} \dot{\eta} dt \cong s(t+T)$, where I have adopted $s(0)$ as the initial condition for $\eta(0)$.  So, basically,
the integral of the auxiliary variable provides an oscillatory estimation for $s(t+T)$. Notice that the subtraction of variables ($\tilde{x}(t+T)-\tilde{x}(t)$) per se already filters noise, since adding random variables effectively decreases their standard deviation. The integral of the auxiliary dynamics contributes further to the noise filtering, 
since the contribution of the noise to the integral can vanish. 

The transmitted binary message is estimated by noticing that 
if $\eta(t)$ is an estimation for $s(t+T)$, and if $y(t)$ is set to be equal to $x(t+2T)$, then, by integrating the inverse dynamics for 1 period, we obtain 
$y(t+T)$, which should produce an estimation of 
$x(t+T)$, and so, the binary encoding $S(t+T)$ of $y(t+T)$, 
should be equal to the binary encoding $s(t+T)$ of $x(t+T)$.

Based on the previous arguments, I redefine $\dot{\eta}$ by a 1-period shift, and use 
\begin{equation}
    \dot{\eta}(t) = {x}(t) - {x}(t-T), 
\label{first-order-filter-murilo}
\end{equation}
\noindent
where the following analysis will assume that there is no noise, and ${x}(t)$ in Eq. (\ref{first-order-filter-murilo}) is the 
trajectory produced by Eq. (\ref{rescaled-corron}). Notice 
that using the auxiliary variable as defined in Eq. (\ref{first-order-filter-murilo}) in the inverse dynamics of Eq. (\ref{rescaled-corron-inverse}), instead of using as in Eq. (\ref{first-order-filter}), only shifts the time of the variable $y(t)$. But this time shift ensures that 
$S(t+T)$ (the encoding of $y(t+T)$) is equal to 
$s(t)$ (the encoding of $x(t)$), being that the $\eta(t)$ is an estimation for $s(t)$.

\begin{figure}[hbt]
\centering
{\includegraphics[scale=0.3]{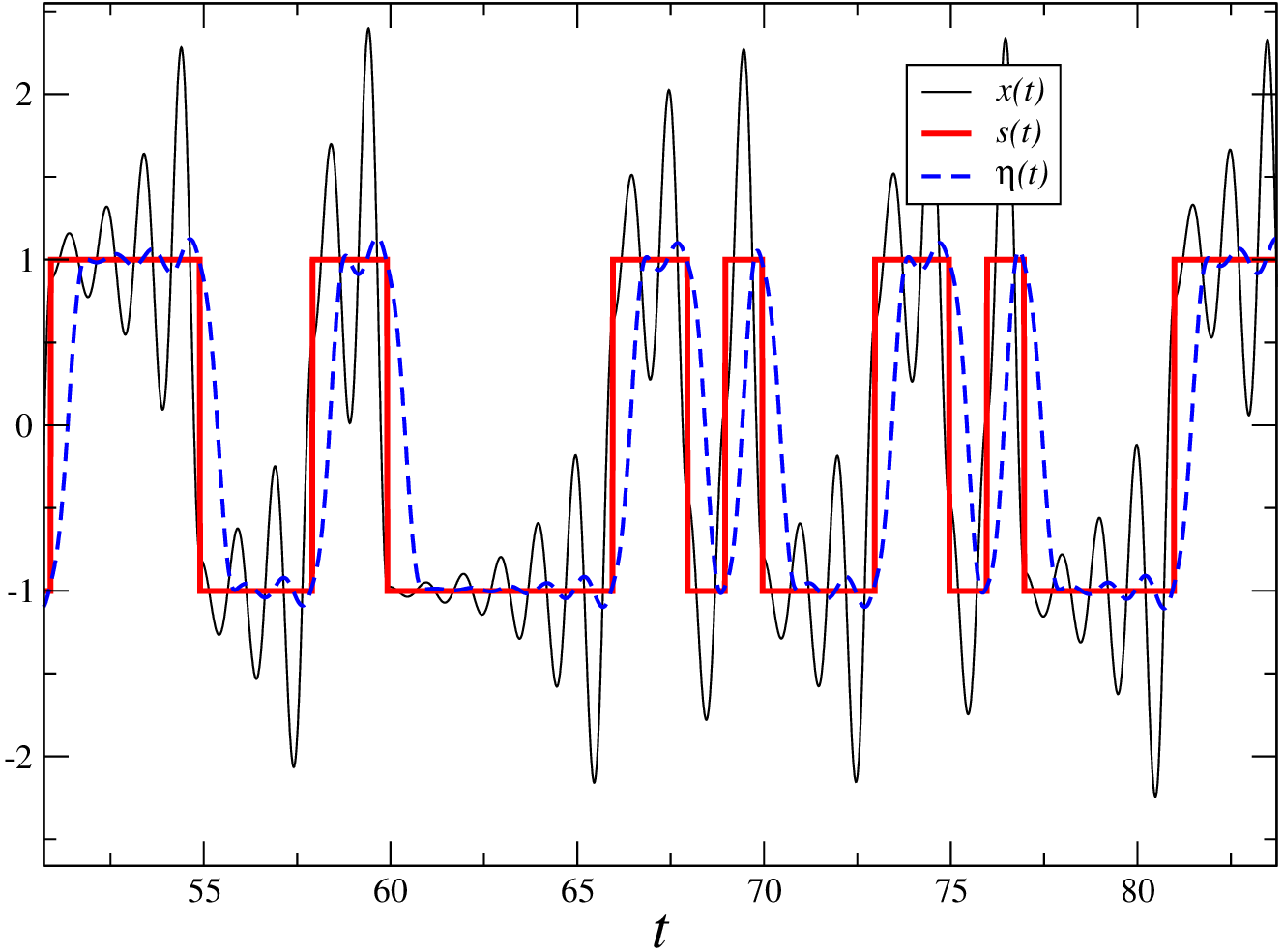}}
\caption{Solutions of variables in the forward dynamics of Eqs. (\ref{rescaled-corron}) and (\ref{rescaled-corron-inverse}) for 
$\gamma=1$ and $\dot{\eta}(t)$ as in Eq. 
(\ref{first-order-filter-murilo}). 
}
\label{sub-fig1}
\end{figure}

Figure \ref{sub-fig1} shows that if $\dot{\eta}(t)$ 
is defined as in Eq. (\ref{first-order-filter-murilo}), 
then $\eta(t)$ (dashed blue line) 
is a good estimation for $s(t)$ (red line), and not of 
$s(t+T)$ if Eq. (\ref{first-order-filter}) had been used. In fact, $\eta(t)$ is time-forwarded estimation of $s(t)$. 
The variable $x(t)$ of the forward dynamics is shown in black line. 

\begin{figure}[hbt]
\centering
{\includegraphics[scale=0.35]{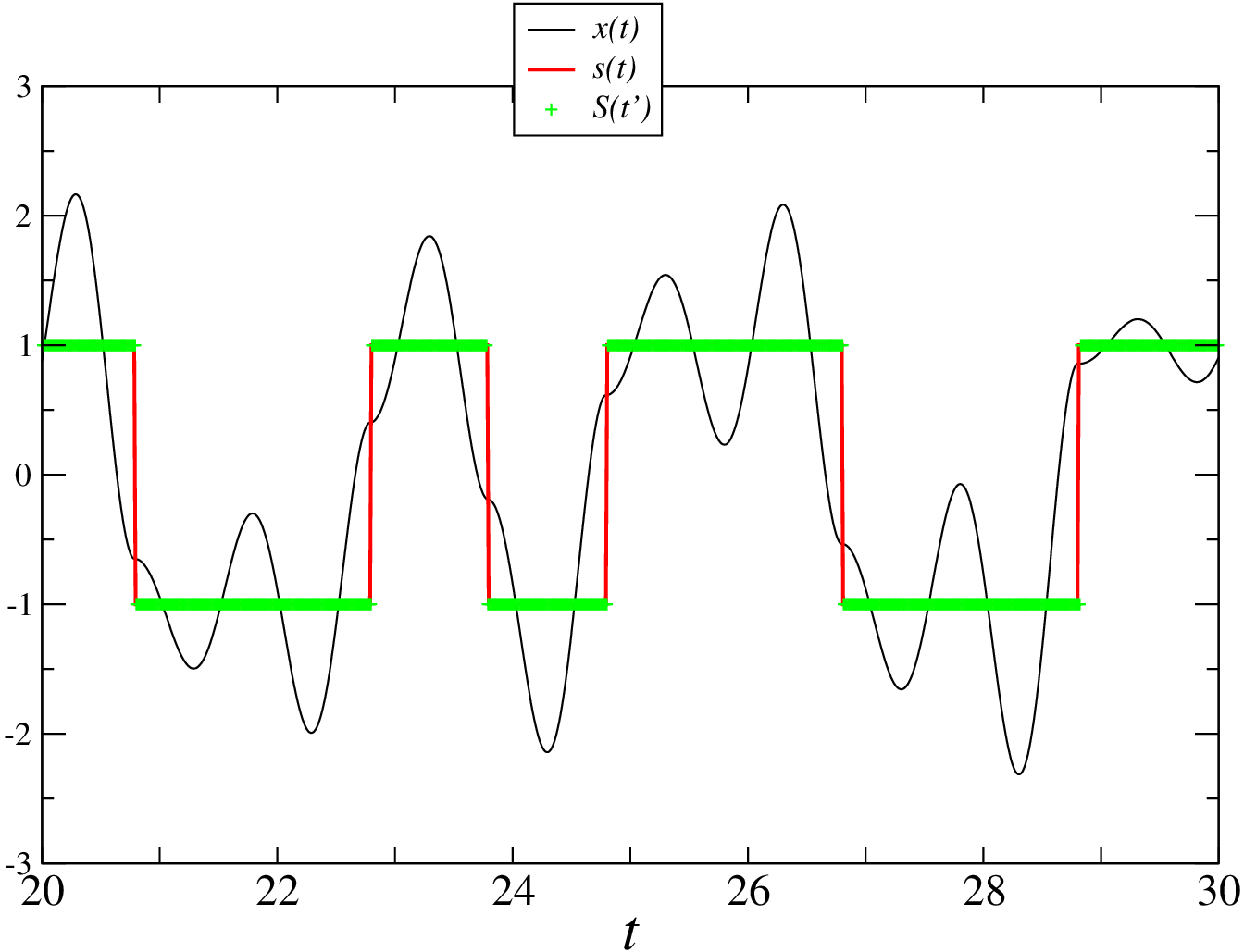}}
\caption{Solutions of variables in the forward dynamics of Eqs. (\ref{rescaled-corron}), $x(t)$ and $s(t)$, 
in black and red lines respectively, and the hybrid symbol $S(t')$ obtained by (\ref{rescaled-corron-inverse}) for 
$\gamma=1$ and $\dot{\eta}(t)$ as in Eq. 
(\ref{first-order-filter-murilo}). $t'=t+T$ in $S(t')$, so to show that $S(t+T)=s(t)$.  
}
\label{sub-fig2}
\end{figure}

Figure \ref{sub-fig2} demonstrates, for the no-noise scenario, that 
the matched 
filter can successfully decode the transmitted symbol. 
This figure shows the signal of the forward dynamics $x(t)$ and respective hybrid variable $s(t)$ (Eq. (\ref{rescaled-corron})), in black and red lines,  respectively. Plus symbols in green colour indicate the values of the continuous hybrid variable $S(t)$ for the matched filter of 
Eq. (\ref{rescaled-corron-inverse}), with $\eta(t)$ defined by (\ref{first-order-filter-murilo}). 

The fact that the red and green curves superimpose in panels demonstrates that for the no-noise scenario, the discrete symbol $S(t)$ obtained from the encoding of the hybrid dynamics of the matched filter 
is equal to the transmitted symbol, $s(t)$, encoding the hybrid dynamics of the forward dynamics. 




\begin{figure}[hbt]
\centering
{\includegraphics[scale=0.3]{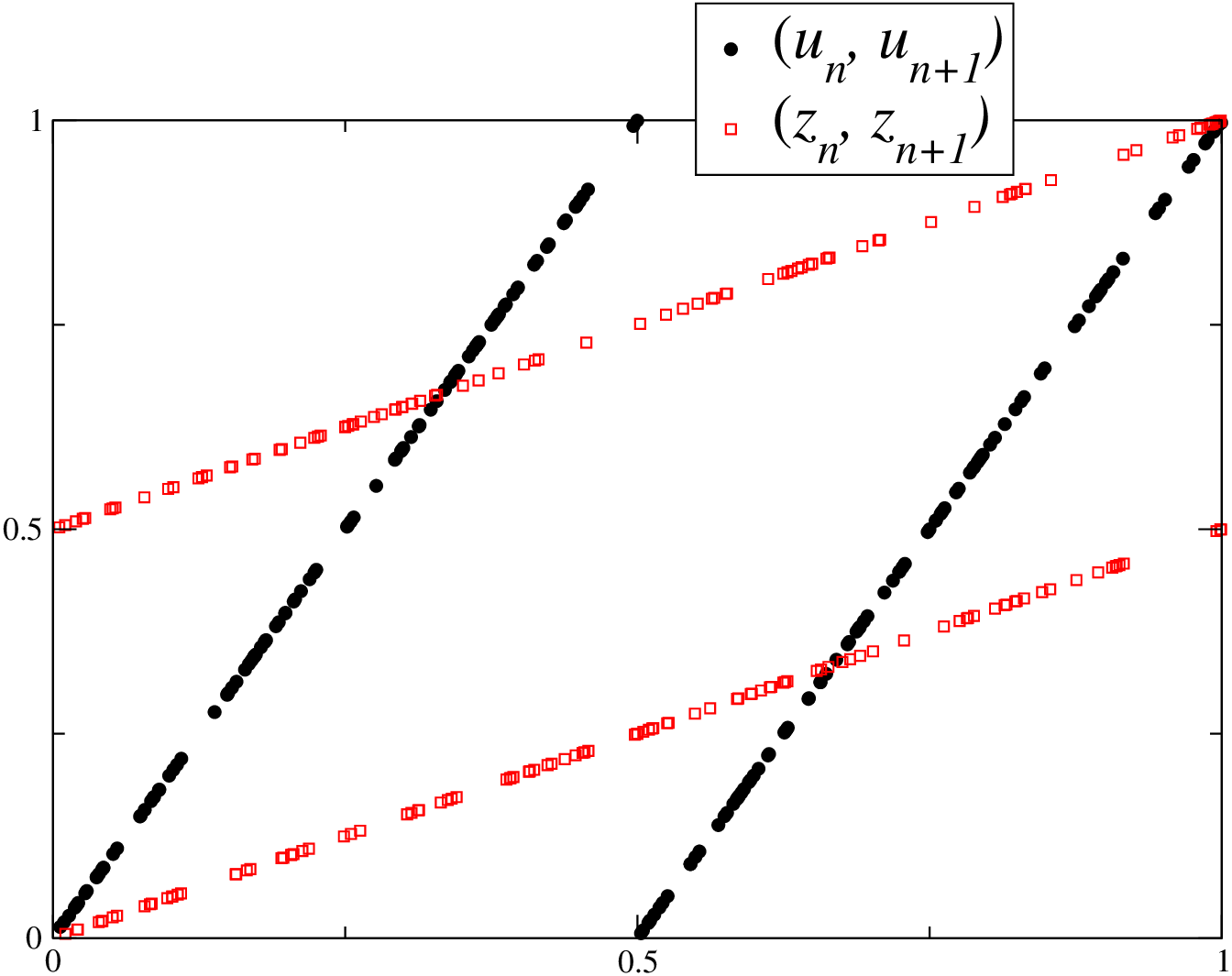}}
\caption{Return mappings of the time-$T$ map of solutions obtained by numerical simulations  of Eq. (\ref{rescaled-corron}), in filled black circles, 
and of Eq. (\ref{rescaled-corron-inverse}), in empty red squares, using that $\eta(t)=s(t)$, $f$=1, and considering only user $k$=1. 
The discrete points were renormalized as usual: 
$u_n = (r_n+1)/2$ and  $z_n = (y_n+1)/2$. }
\label{sub-fig3}
\end{figure}

Let me now argue that 
the system of Eqs. (\ref{rescaled-corron}) and (\ref{rescaled-corron-inverse}), coupled by making the rough  approximation
\begin{equation}
    \eta(t)=s(t),
\end{equation}
\noindent
can be approximately described by the unfolded-Baker's map. 
Then, I make further conclusions about how decoding in the multiuser environment can be done.  As recognized in Ref. 
\cite{bailey2016}, making $\eta(t)=s(t)$ in Eq. (\ref{rescaled-corron-inverse}), makes this equation to describe what was called ``reverse time chaos" in Ref.  \cite{corron2006chaos}.   

Let us take the values of $y$ in Eq. (\ref{rescaled-corron-inverse}) at discrete times $nT$, writing that $y(nT)=y_n$, and define the new variable for users 1 and 2 as before  $y^{(1)}_n=2z^{(1)}_n -1$ and $y^{(2)}_{2n}=2z^{(2)}_n -1$. 

Figure \ref{sub-fig3} shows in black circles and red squares that time-$T$ mappings of solutions of Eq. (\ref{rescaled-corron}) and of Eq. (\ref{rescaled-corron-inverse}) for $\eta(t)=s(t)$, $f$=1, and for user $k$=1 are the mappings presented in  
Eqs. (\ref{user1-shift-SM}) and (\ref{inverse-map}), respectively.  

If Equations  
\begin{eqnarray}
u^{(1)}_{n+1}=2u^{(1)}_n-\lfloor 2u^{(1)}_n \rfloor \equiv 2u^{(1)}_n- b^{(1)}_n, \label{user1-shift-SM} \\
u^{(2)}_{n+1}=4u^{(2)}_n-\lfloor 4u^{(2)}_n \rfloor \equiv 4u^{(2)}_n- b^{(2)}_n, \label{user2-shift-SM} 
\end{eqnarray}
\noindent
are map solutions of Eq. (\ref{rescaled-corron}) (in the re-scaled coordinate system, with appropriate $\gamma$ gains) for user $k$ with frequencies 
$f^{(k)}=k$, their inverse mapping the solution of Eq. (\ref{rescaled-corron-inverse}) is given by 

\begin{equation}
z^{(k)}_{n+1}=2^{-k}\{z^{(k)}_n-\lfloor 2^{k}u^{(k)}_n \rfloor\}, \mbox{\ and } \lfloor 2^{k}u^{(k)}_n \rfloor \equiv b^{(k)}_n.  
\label{inverse-map}
\end{equation}
This map can be derived simply defining $z^{(k)}_{n+1} = u^{(k)}_{n}$ and  
$z^{(k)}_{n} = u^{(k)}_{n+1}$. We always have that $\lfloor 2^{k}u^{(k)}_n  \rfloor = b^{(k)}_n$. So, for any 
$z^{(k)}_n \in [0,1]$ and which can be simply chosen to be equal to the received combined signal $O_n$ (normalized such that $\in [0,1]$), it is also true that  
\begin{equation}
\lfloor 2^{k}    z^{(k)}_{n+1} \rfloor = \lfloor 2^{k}u^{(k)}_n  \rfloor = b^{(k)}_n.  
\label{decoding-by-variable}
\end{equation}
\noindent
So, if we represent an estimation of the transmitted symbol of user $k$ by $\tilde{b}^{(k)}_n$, then decoding of the transmitted symbol of user $k$ can be done by calculating $z^{(k)}_{n+1}$ using the inverse dynamics of the user $k$
\begin{equation}
z^{(k)}_{n+1}=2^{-k}\{z^{(k)}_n - \tilde{b}^{(k)}_n\}.   
\label{inverse-map-decode3}
\end{equation}
\noindent
and applying this value to Eq. (\ref{decoding-by-variable}).

Initial conditions in Eqs. (\ref{inverse-map-decode3}) need to be set according to the decoding scheme and the uncertainty of the received signal. As long as $\tilde{b}^{(k)}_n\ \in (0,1)$ (so, can assume either the value 0 or 1), then regardless of $z^{(k)}_n \in [0,1]$, one will obtain 
that $\lfloor 2^{k}u^{(k)}_n  \rfloor = \tilde{b}^{(k)}_n$. So, if 
a symbol $b^{(k)}_n \in (0,1)$ is wrongly estimated, the use of Eq. (\ref{decoding-by-variable}) cannot correct that. However, $\tilde{b}^{(k)}_n \in [0,1/2[$ (or $\tilde{b}^{(k)}_n \in [1/2,1]$) can by the use of 
Eqs. (\ref{inverse-map-decode3}) and with the decoding rule as in Eq. (\ref{decoding-by-variable}) correctly estimate the value of $b^{(k)}_n$. The filter in \cite{corron2010matched} works based on this dynamical principle, in which the inverse dynamics is forced by the auxiliary variable, assumed to provide a good estimation of the transmitted symbol, with a shift in time depending on how $\dot{\eta}$ is defined. 

The result in Eq. (\ref{inverse-map}) for $k=1$ can also be derived noticing that as shown 
in Ref. \cite{corron2006chaos}, Eq. (\ref{rescaled-corron-inverse}) after the proper rescaling also done in here for the variables and symbolic sequence has a solution equal to $z^{(1)}_{n} =  \sum_{j=1}^{\infty} 2^{-j} b^{(1)}_{n-j}$. Similarly, a solution in terms of the symbols for the variable $u_n$ in Eq. (\ref{user1-shift-SM}) can be written as 
$u^{(1)}_{n} = \sum_{j=0}^{\infty} 2^{-(j+1)} b^{(1)}_{j}$. This means that the system formed by the variables 
$u^{(k)}_{n},z^{(k)}_{n}$ is a generalization (for $k \neq 1$)  of the unfolded Baker's map \cite{lasota1985probabilistic}, 
being described by a time-forward 
variable $u^{(k)}_{n}$ (the Bernoulli shift for $k$=1), and its backward  variable component $z^{(k)}_{n}$.   

\section{The negative Lyapunov Exponents of the inverse dynamics}

To time reverse the hybrid system in Eq. (\ref{rescaled-corron}), we apply the time-rescaling $d/dt^{\prime} = -d/dt$, which basically produces a system of ODEs  that is the one in Eq. (\ref{rescaled-corron}) for a time-reverse.  This system of ODEs is the one 
in Eq. (\ref{rescaled-corron-inverse}).


Such time-rescaling simply reverts the signs of the spectrum of LEs of the time-forward dynamics \cite{eckmann1985ergodic}, thus it should possess a pair of degenerated exponents with a value of  
\begin{equation}
\chi = - \beta.     
\label{1d-le-negative}
\end{equation}

\section{Non-orthogonal multiple access (NOMA)}

The understanding of how to decompose periodic signals with equal or different frequencies combined together with appropriate power gains for each signal for communication purposes is not new \cite{kizilirmak2016non,benjebbour2017overview,dai2018survey}. And it is in fact a promising approach to deliver the demand required for 5G systems.  To cope with the expected demand in 5G wireless communication, non-orthogonal multiple access (NOMA) \cite{kizilirmak2016non,benjebbour2017overview,dai2018survey} was proposed to allow all users to use the whole available frequency spectrum. One of the most popular NOMA scheme allocate different power gains to the signal of each user. 

Suppose $y_k(t)$ represents an orthogonal frequency-division multiplexing (OFDM) periodic wavesignal for an user $k$ \cite{hampton2013introduction}
\begin{eqnarray}
y_k(t)=\Re \left\{ \sum_{k=0}^{N_s-1}s(k)e^{j2\pi(f_c + k/T_{OFDM})t} \right\},\\
\mbox{\ } 0 \leq t \leq T_{OFDM},     \nonumber 
\label{OFDM-signal}
\end{eqnarray}
\noindent
where $T_{OFDM}\equiv N_sT_s$ represents the total length of the input signal, $N_s$ the number of symbols to be transmitted by user $k$, and $s(k)=s(kT_s)$, with $k=0,\ldots,N_s-1$ are the symbols to be transmitted (in the frequency domain, so basically representing a discrete alphabet).  Equation (\ref{OFDM-signal}) is nothing but the inverse Fourier transformation of the symbols.
The user uses subcarrier frequency bands spaced by $1/T_{OFDM}$ and centered at $(f_c + k/(2T_{OFDM}))$. Orthogonality of the signal between any two non-equal subcarriers means that $\sum_{k=0}^{N_s-1} e^{j2\pi(k_1/T_{OFDM})t} e^{j2\pi(k_2/T_{OFDM})t}=0$, for $k_1 \neq k_2$. In the OFDM communication scheme, each user must use a different set of subcarriers. For example, user 1 communicates with $f_c=1/T_{OFDM}$ and user 2 communicates with $f_c=N_s/T_{OFDM}$. For user 1, regardless of the subcarrier the bit rate is dominated by the lowest frequency. Moreover, users cannot use the same set of subcarriers, thus limiting the efficient use of the spectrum resources. 


For the uplink scenario, this would translate into a signal being received in BS described by 
\begin{equation}
O(t)_{up}=\sum_{k=1}^N \sqrt{\gamma^{(k)}} g_k y_k(t) + w(t), 
\label{NOMA-uplink}
\end{equation}
\noindent
where $\gamma^{(k)}$ is the average power of user $k$ signal, $g_k$ represents the attenuation for the shortest path link between the BS and the user $k$, and $w(t)$ is AWGN. The signal $y_k(t)$ is the one transmitted by user $k$. Other signal not propagating along the shortest path are treated as noise. 

For the downlink communication, when a BS sends 1 signal to several users, the signal transmitted by the station is described by 
\begin{equation}
O(t)_{down}=\sum_{k=1}^N \sqrt{\gamma^{(k)}} y_k(t), 
\label{NOMA-downlink-transmitted}
\end{equation}
\noindent
and the signal received by user $k$ is described by 
\begin{equation}
O^{(k)}(t)_{down}=  g_k O(t)_{down} + w_k(t)
\label{NOMA-downlink-received}
\end{equation}
\noindent
where $g_k$ is the attenuation factor between the BS and the user $k$ along the shortest path propagation.

Notice that considering only the direct path ($L=1$), equations in the main manuscript modelling the uplink and downlink scenarios are equivalent to Eqs. (\ref{NOMA-uplink}), (\ref{NOMA-downlink-transmitted}), and (\ref{NOMA-downlink-received}). 
There are however crucial differences between the Wi-C1 being here proposed and the NOMA. 
Each user in the Wi-C1 has a unique natural central frequency $f$, however 
$x(t)$ is a chaotic signal and as such it is naturally 
broadband. Signals from different users will always interfere. In the OFDM-based NOMA scheme, bits are "slowed" modulated. Each subcarrier modulates a symbol per channel use ($n T_s \leq t \leq (n+1) T_s$), whereas in the Wi-C1, a user modulates $f$ digital 
symbols per channel use, or $f$bits per channel use ($nN_s \leq t \leq (n+1) N_s$). In NOMA, encoding as well as decoding employs Fourier functions. In the Wi-C1, decoding can be trivially done, or using simple filters. Finally, nothing impedes Wi-C1 to have users operating with arbitrary central frequencies and their multiples, using simultaneously the whole spectrum. This however is out of the scope of the present work.

\bibliographystyle{apsrev4-2} 
\bibliography{apssamp}